\documentclass[aps,numerical,superscriptaddress,floatfix,reprint,groupedaddres,longbibliography]{revtex4-1}

\usepackage{graphicx,color}
\usepackage{amsmath,amssymb,amsfonts,bm}
\usepackage[colorlinks, linkcolor=red,anchorcolor=green,citecolor=blue]{hyperref}

\newcommand{\be}{\begin{equation}}
\newcommand{\ee}{\end{equation}}
\newcommand{\ba}{\begin{eqnarray}}
\newcommand{\ea}{\end{eqnarray}}
\newcommand{\ban}{\begin{eqnarray*}}
\newcommand{\ean}{\end{eqnarray*}}

\def\v2{\mbox{$v_2$}}

\usepackage{multirow}

\begin{document}

\title{Implications of the isobar run results\\ for chiral magnetic effect in heavy ion collisions}
\medskip

\author{Dmitri E. Kharzeev}
\email{dmitri.kharzeev@stonybrook.edu}
\affiliation{ Center for Nuclear Theory, Department of Physics and Astronomy, Stony Brook University, Stony Brook, New York 11794-3800, USA}
\affiliation{Department of Physics, Brookhaven National Laboratory, Upton, New York 11973-5000, USA}

 \author{Jinfeng Liao}
\email{liaoji@indiana.edu}
\affiliation{ Physics Department and Center for Exploration of Energy and Matter, Indiana University, 2401 N Milo B. Sampson Lane, Bloomington, IN 47408, USA}

\author{Shuzhe Shi}
\email{shuzhe.shi@stonybrook.edu}
\affiliation{ Center for Nuclear Theory, Department of Physics and Astronomy, Stony Brook University, Stony Brook, New York 11794-3800, USA}

\date{\today}

\begin{abstract}
Chiral magnetic effect (CME) is a macroscopic transport phenomenon induced by quantum anomaly in the presence of chiral imbalance and an external magnetic field.  Relativistic heavy ion collisions provide the unique opportunity to look for CME in a non-Abelian plasma, where the chiral imbalance is created by topological transitions similar to those occurring in the Early Universe. 
The isobar run at Relativistic Heavy Ion Collider was proposed as a way to separate the possible CME signal driven by magnetic field from the background. The first blind analysis results from this important experiment have been recently released  by the STAR Collaboration. Under the pre-defined assumption of identical background in RuRu and ZrZr, the results are inconsistent with the presence of CME, as well as with all existing theoretical models (whether including CME or not). However the observed difference of backgrounds must be taken into account before any physical conclusion is drawn. 
In this paper, we show that once the observed difference in hadron multiplicity and collective flow are quantitatively taken into account, the STAR results could be  consistent with a finite CME signal contribution of about $(6.8\pm2.6)\%$.
\end{abstract}

\maketitle

{\em Introduction.---}  Chiral magnetic effect (CME) is a macroscopic transport phenomenon induced by quantum anomaly in chiral matter.  In the presence of an external magnetic field $\mathbf{B}$ and a chiral imbalance, the CME amounts to the generation of an electric current $\mathbf{J}$ along $\mathbf{B}$: 
\begin{eqnarray}
\mathbf{J} = \sigma_5 \mathbf{ B} ,
\end{eqnarray}
where   $\sigma_5$ is the chiral magnetic conductivity that is proportional to the chiral chemical potential parameterizing the chirality imbalance between the left- and right-handed chiral fermions~\cite{Kharzeev:2004ey,Kharzeev:2007jp,Fukushima:2008xe}.  The CME has an impact on the physics of high-density QCD matter~\cite{Kharzeev:1998kz,Kharzeev:2001ev,Kharzeev:2004ey,Kharzeev:2007tn,Kharzeev:2007jp,Fukushima:2008xe}, condensed matter physics~\cite{Son:2012bg,Zyuzin:2012tv,Basar:2013iaa,Li:2014bha,Huang:2015eia}, astrophysics~\cite{Grabowska:2014efa,Masada:2018swb,Yamamoto:2015gzz}, cosmology~\cite{Tashiro:2012mf,Vilenkin:1982pn,Vilenkin:1980fu}, 
plasma physics~\cite{,Akamatsu:2013pjd,Hirono:2016jps,Gorbar:2016qfh}, and quantum information~\cite{Kharzeev:2019ceh,Shevchenko:2012nv}; 
for reviews, see e.g.~\cite{Kharzeev:2020jxw,Kharzeev:2015znc,Kharzeev:2013ffa,Fukushima:2018grm,Hattori:2016emy,Gao:2020vbh,Burkov:2015hba,Armitage:2017cjs,Shovkovy:2021yyw}. 
  
Relativistic heavy ion collisions provide a unique opportunity to create and study a quark-gluon plasma (QGP) at a temperature of over a trillion degrees. 
In QGP, the fluctuations of quark chirality imbalance are generated through the  topological fluctuations of gluon fields. Moreover, the QGP produced in heavy ion collisions experiences an extremely strong magnetic field~\cite{Kharzeev:2007jp}  created mostly by the fast-moving spectator protons. Thus the CME is expected to occur in the produced QGP~\cite{Kharzeev:2004ey}, and may lead to a detectable signal in these collisions~\cite{Voloshin:2004vk}. The observation of CME in heavy ion collisions would establish the presence of topological fluctuations in a non-Abelian plasma, which represent a crucial ingredient of the baryon asymmetry generation in the Early Universe. 

Extensive experimental efforts have been made by STAR, ALICE and CMS Collaborations to look for CME in collisions at both the Relativistic Heavy Ion Collider (RHIC) and the Large Hadron Collider (LHC)~\cite{STAR:2009wot,STAR:2009tro,ALICE:2020siw,ALICE:2017sss,CMS:2016wfo}, see reviews~\cite{Zhao:2019hta,Li:2020dwr,Bzdak:2019pkr}. The search has proved to be challenging due to a relatively small signal masked by a strong background contamination~\cite{Voloshin:2004vk,Wang:2009kd,Pratt:2010gy,Bzdak:2009fc,Bzdak:2010fd}, see e.g. discussions in~\cite{Zhao:2019hta,Li:2020dwr,Bzdak:2019pkr,Bzdak:2012ia,Choudhury:2021jwd,Christakoglou:2021nhe}. 

To disentangle the signal driven by magnetic field (in addition to topological fluctuations) and the background driven by the collective flow determined by the collision geometry, it has been proposed to perform a measurement of CME observables in RuRu and ZrZr isobar collisions~\cite{Voloshin:2010ut,Skokov:2016yrj}. The motivation for this measurement was that the similar size and shape of the colliding nuclei would lead to a nearly identical background, whereas the difference in electric charge of Ru and Zr nuclei would result in a difference in the created magnetic field, and thus in a difference in the observed CME signal.

In 2018 the STAR Collaboration performed the corresponding measurements at RHIC. A careful blind analysis was carried out subsequently~\cite{STAR:2019bjg}, with the first set of data released in September 2021~\cite{STAR:2021mii}. STAR results are inconsistent with the ``pre-defined" criteria for the CME, i.e. the criteria based on the assumption that the backgrounds in RuRu and ZrZr collisions are identical. Namely, the ratios of the CME observables measured in RuRu and ZrZr collisions are smaller than 1, whereas a stronger magnetic field in RuRu system would apparently make this ratio bigger than 1 in the presence of CME.
The problem with this result however is that if the CME is absent, the ratios of these observables would have to be {\it equal} to 1, and not be smaller than 1. Indeed, none of the theory models  predicted the ration smaller than 1, so this experimental result begs for an explanation. 

The examination of STAR data~\cite{STAR:2021mii} shows the key to understanding this puzzle is the observed difference between the gross properties of hadron production in RuRu and ZrZr collisions that stem from the difference in the shape and size of Ru and Zr nuclei. This {\it observed} difference in the multiplicity distributions and the collective flow invalidates the ``pre-defined" criteria for the presence of CME, and clearly indicates the need for a post-blinding re-analysis of STAR data. Only after such an analysis is performed, one will be able to draw conclusions about the presence or absence of CME in the data.
In this Letter, we address this issue by combining insights from theoretical simulations based on the event-by-event anomalous-viscous fluid dynamics (EBE-AVFD) framework~\cite{Shi:2019wzi,Shi:2017cpu,Jiang:2016wve,An:2021wof,Shen:2014vra} with the analysis of STAR data.
\vskip0.3cm

{\em Correlation observables.---}  In heavy ion collisions, the CME leads to a charge separation along the magnetic field which is approximately perpendicular to the reaction plane~\cite{Bloczynski:2012en}. Such a charge separation can be measured via charge-dependent azimuthal correlations~\cite{Voloshin:2004vk,STAR:2009wot,STAR:2009tro}, with the most commonly used $\Delta \gamma$ and $\Delta \delta$ observables defined as: 
\begin{eqnarray} 
&& \Delta \gamma \equiv  \left[  \cos (\phi_1 + \phi_2 - 2\Psi_2 ) \right]_{OS-SS}    \ , \\
&&  \Delta \delta  \equiv   \left[  \cos (\phi_1 - \phi_2  ) \right]_{OS-SS}  \ . 
\end{eqnarray}
In the above, $\phi_{1,2}$ are azimuthal angles of the charged hadron pairs while $\Psi_2$ is the event-plane angle. The ``OS-SS'' means the difference between the opposite-sign hadron pairs (i.e. the pairs of hadrons with opposite electric charges) and same-sign pairs.  Other observables have also been developed and used for experimental analysis~\cite{STAR:2021mii,Choudhury:2021jwd}, such as $\Delta \gamma$ comparison between reaction and event plane~\cite{Xu:2017qfs,Voloshin:2018qsm}, $\Delta \gamma$ invariant mass dependence~\cite{Zhao:2017nfq}, R-correlator~\cite{Magdy:2017yje}, signed balance function~\cite{Tang:2019pbl}, event-shape engineering~\cite{ALICE:2017sss,Wen:2016zic} and others.

The CME signal induces the parity-odd harmonic in the azimuthal angle distribution of charged hadrons~\cite{Kharzeev:2004ey}:
$$
\frac{dN_\pm}{d \phi} \sim 1 \pm 2 a_1 \sin (\phi - \Psi_2) + ...
$$
Therefore it contributes to the above observables as $\Delta \gamma_{cme} = + 2 a_1^2 $ and  $\Delta \delta_{cme} = - 2 a_1^2 $ which are thus proportional to the square of the magnetic field strength. 

The main challenge in the experimental search of CME is the background correlations that dominate the observables. The identified backgrounds  are local charge conservation at hydrodynamic freeze-out and resonance decays. Their contributions to the observables  scale approximately as $\Delta \gamma_{bkg} \propto  + \frac{v_2}{N_{ch}} $ and  $\Delta \delta_{bkg} \propto  + \frac{1}{N_{ch}}$ where $N_{ch}$ is the charged particle multiplicity and $v_2$ is the elliptic flow coefficient. One may also consider the observable $\Delta {\bar \gamma} \equiv \Delta \gamma/v_2$ for which $\Delta {\bar \gamma}_{bkg} \propto  + \frac{1}{N_{ch}}$. Such scaling behaviors were found to approximately hold in model simulations. For more detailed discussions on signals and backgrounds see e.g.~\cite{Choudhury:2021jwd}. 

The isobar collision experiment in principle allows to separate the signal and background contributions.  In the idealized scenario, the two systems would be identical in their bulk properties (such as multiplicity and collective flow), which would result in the identical background contributions to the $\Delta \gamma$ and $\Delta \delta$ correlators. 
Therefore in the case of pure background, with no CME present, the measured isobar ratios would be 
$ \Delta \gamma^{Ru}/  \Delta \gamma^{Zr} =1$ and $ \Delta \delta^{Ru}/  \Delta \delta^{Zr} =1$.  The case of a finite CME contribution would imply  $ \Delta \gamma^{Ru}/  \Delta \gamma^{Zr} > 1$ and $ \Delta \delta^{Ru}/  \Delta \delta^{Zr} < 1$. These are the ``pre-defined criteria" used in the STAR blind analysis~\cite{STAR:2021mii,STAR:2019bjg}. However, as clearly shown by the STAR data, the bulk properties of hadrons produced in the two isobar systems are {\em not} identical. For example, in the same centrality class, the hadron multiplicities differ at a few percent level -- this difference is extremely important in the search for the $\sim 1\%$ CME effect. This situation requires a more careful isobar comparison with a proper baseline identification.

\begin{figure}[!hbt]
	\begin{center}
		\includegraphics[width=2.5in]{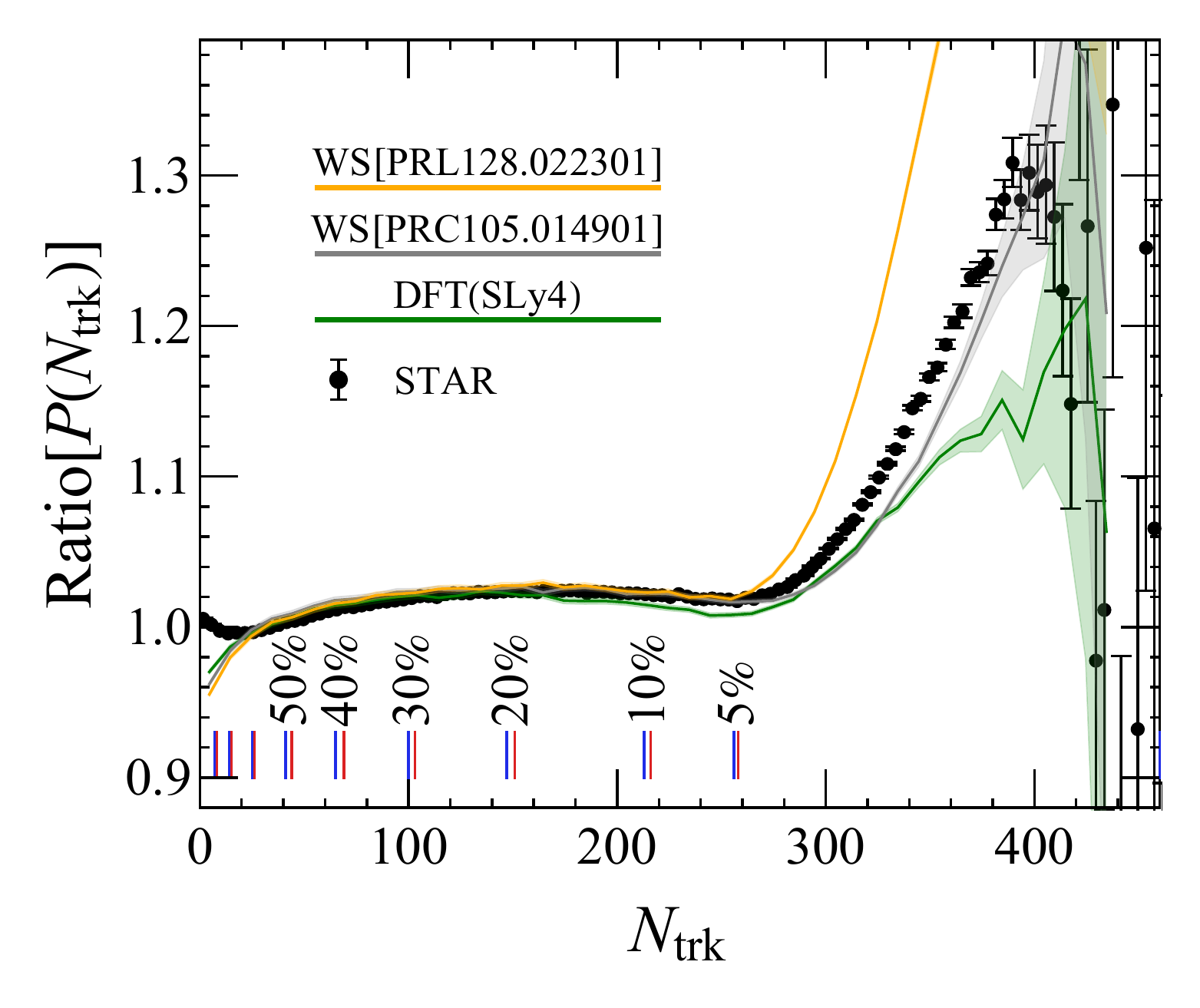}
		\includegraphics[width=2.5in]{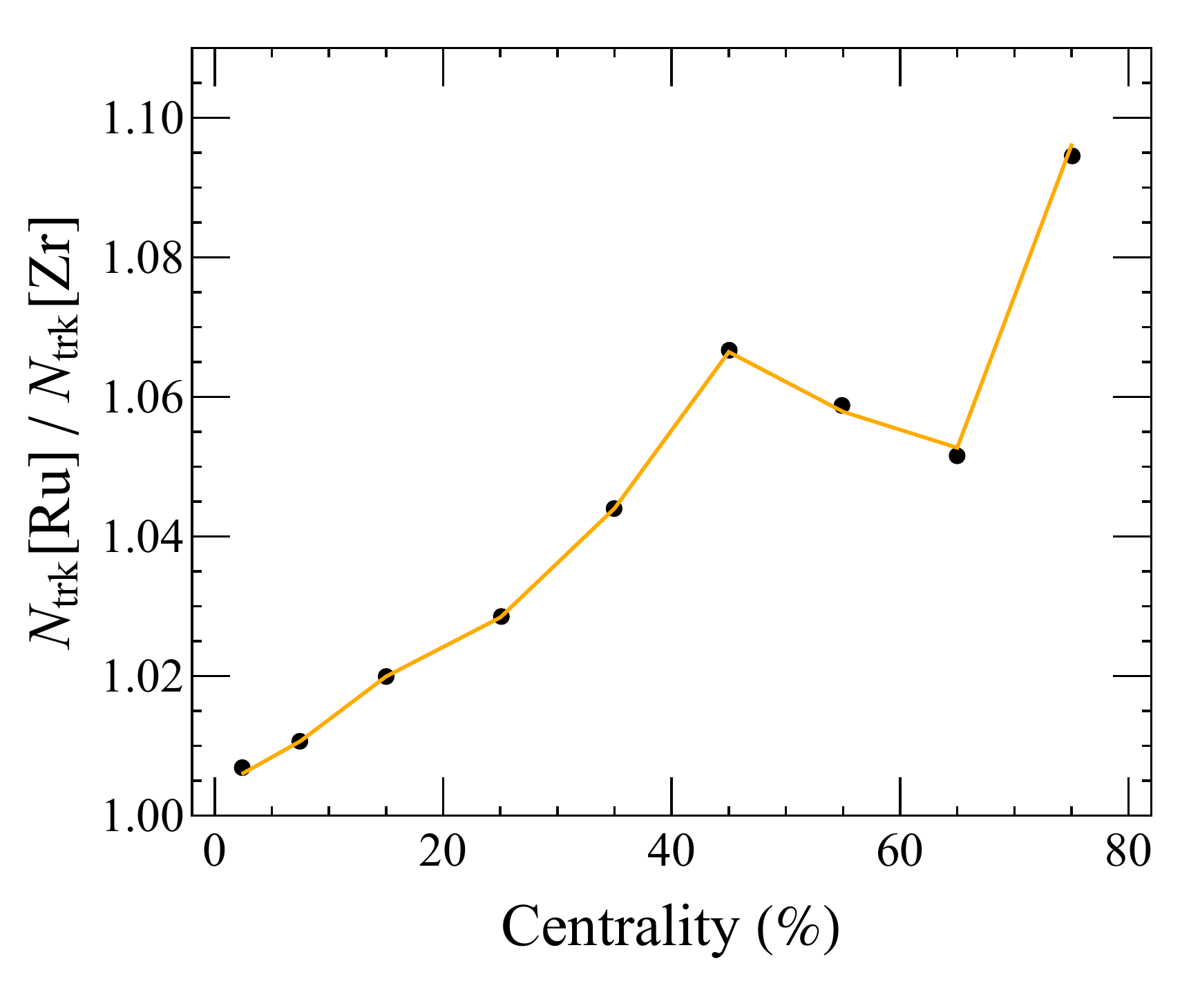}
		\caption{The Ratios of the multiplicity distribution $P(N_{ch})$ (upper panel) and of the average multiplicity  in the same centrality class measured in RuRu and ZrZr collisions as defined by the STAR analysis (lower panel). The orange, grey and green curves are simulation results taking initial nucleon distributions in the colliding nuclei to be deformed~\cite{Zhang:2021kxj} and spherical~\cite{STAR:2021mii} Woods-Saxon distribution and Density Functional Theory calculation~\cite{Xu:2021vpn}, respectively. In the upper panel, red(blue) vertical bars indicate centrality class definition by STAR analysis for RuRu(ZrZr) collisions.}
		\vspace{-0.5cm}
		\label{fig_1}
	\end{center}
\end{figure}
\vskip0.3cm

{\em Isobar  multiplicity comparison.---}  As discussed above, the event multiplicity plays a key role in the background correlations and it is important to first examine the multiplicity difference between the two isobar pairs. While the Ru and Zr nuclei have an equal number of nucleons, the geometric distributions of protons and neutrons within these nuclei have a  non-negligible difference~\cite{Shi:2018sah,Xu:2017zcn,Hammelmann:2019vwd}. This difference translates into the difference in the initial conditions (e.g. the participant and the binary collision densities), which in turn affects the subsequent bulk evolution and leads to the observed discrepancy in multiplicity. 

First, we will show that by adopting suitable parameters (like charge radius, neuron skin and harmonic deformation coefficients) for Ru and Zr nuclear distributions, one can reasonably reproduce the observed multiplicity difference. In Fig.~\ref{fig_1} (upper panel), we show the ratio for the multiplicity distribution $P(N_{ch})$ between RuRu and ZrZr from simulations with several choices for the nuclear parameters~\cite{Zhang:2021kxj,STAR:2021mii,Xu:2021vpn}.  The STAR data are also shown for comparison, along with vertical bars indicating centrality class definition by STAR analysis. The simulation results compare well with  data for the $(20\sim 50)\%$ centrality class which will be the focus of our analysis. Considerable deviations occur in the very central and peripheral  regions where fluctuations and uncertainties of both simulations and data become large. In Fig.~\ref{fig_1} (lower panel) for the ratio of average multiplicity between RuRu and ZrZr in the same centrality class as defined by the STAR analysis, one sees nice agreement between simulation results and experimental data. With such multiplicity difference quantitatively accounted, one can expect the simulations to provide useful and realistic baseline resulting from the background correlations.

\begin{figure*}[t]
	\begin{center}
		\includegraphics[width=2.5in]{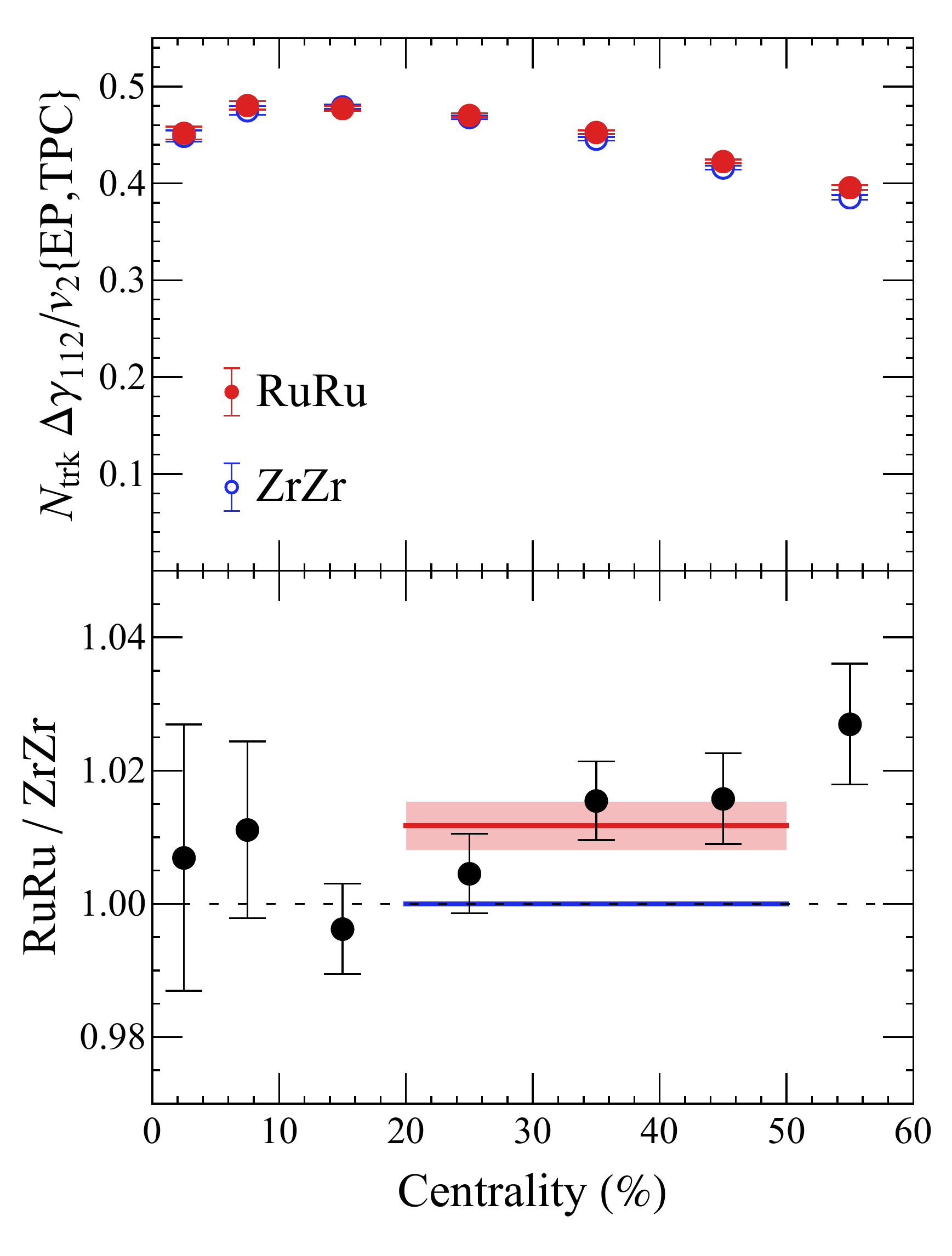}  \hspace{0.8cm}
		\includegraphics[width=2.5in]{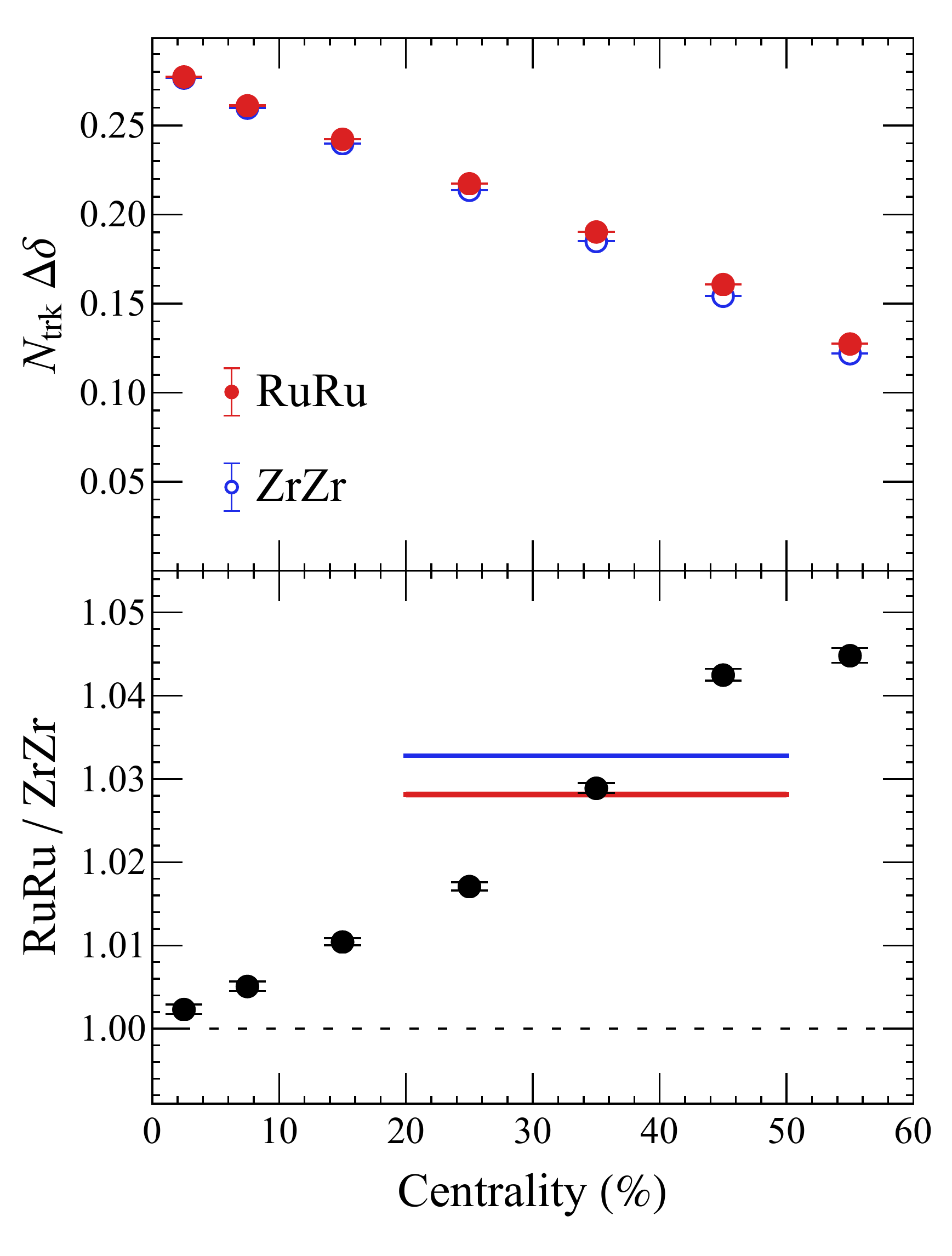} 
		\caption{Comparison between RuRu and ZrZr measurements for scaled correlators $N_{ch} \times \Delta \gamma / v_2$ (left) and $N_{ch} \times \Delta \delta$ (right), with the lower panels showing the RuRu to ZrZr ratios. The STAR data are taken from~\cite{STAR:2021mii,hepdata.115993}. In the lower panels, the red shaded bands indicate measured values with error bars for $(20\sim 50)\%$ while the blue lines indicate the baselines from the present simulation analysis.}
		\vspace{-0.5cm}
		\label{fig_2}
	\end{center}
\end{figure*}

\begin{figure}[!hbt]
	\begin{center}
		\includegraphics[width=2.5in]{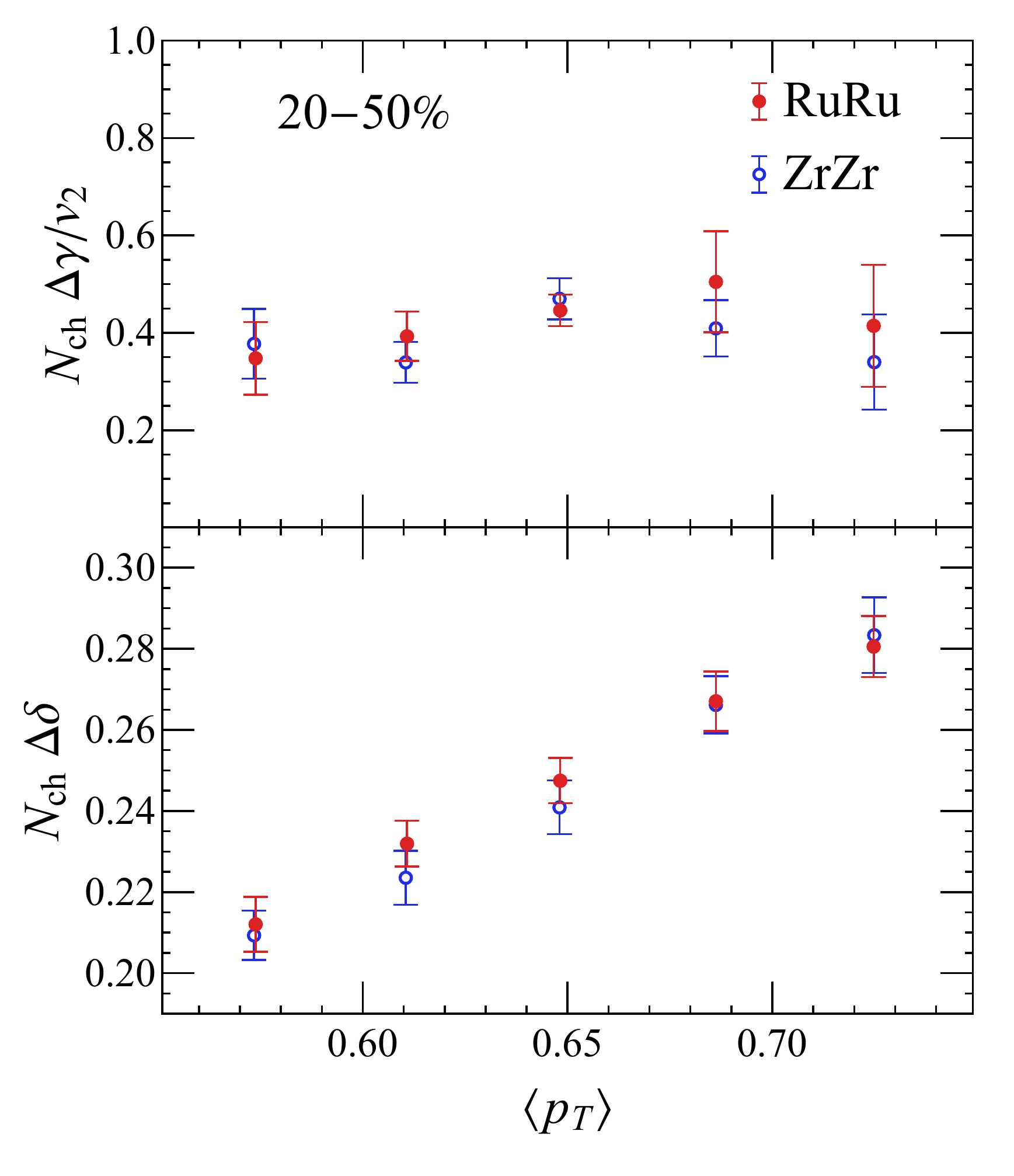}
		\caption{The dependence of scaled $\gamma$ (upper) and $\delta$ (lower) correlators from pure  background contributions on the  average transverse momentum $\langle p_T \rangle$ from  simulations for $(20\sim 50)\%$ centrality.   }
		\vspace{-0.5cm}
		\label{fig_3}
	\end{center}
\end{figure}
 
 {\em Understanding the measured correlations.---} Given the isobar multiplicity difference, a reasonable way to compare the correlators is to take into account   the expected scaling behavior  of background correlations by examining the following re-scaled correlators: $N_{ch} \times \Delta \gamma / v_2$ and $N_{ch} \times \Delta \delta$. Using STAR data from~\cite{STAR:2021mii,hepdata.115993},  we plot these re-scaled observables  for RuRu and ZrZr in Fig.~\ref{fig_2} (upper panels) as well as the RuRu/ZrZr ratios (lower panels).  
 
Let us discuss the RuRu/ZrZr ratios shown in the lower panels of Fig.~\ref{fig_2}. If the background correlations were to scale exactly as $v_2/N_{ch}$, then the pure-background baseline for both $N_{ch} \times \Delta \gamma / v_2$ and $N_{ch} \times \Delta \delta$ ratios would be unity. (As a caveat, the non-flow effect has the potential of shifting this ratio at about $1\%$ level and requires careful scrutiny~\cite{Feng:2021pgf}.)
A nonzero CME contribution on top of the baseline would then lead to $R_{\gamma}\equiv [N_{ch} \times \Delta \gamma / v_2 ]_{RuRu/ZrZr} > 1$ and $ R_{\delta}\equiv [N_{ch} \times \Delta \delta ]_{RuRu/ZrZr}  < 1$. As one can see from Fig.~\ref{fig_2}, the ratio for the scaled $\gamma$ correlator is around unity for very central collisions and gradually increases to a value well above one  towards the relatively more peripheral region. 

This trend is consistent with a nonzero CME contribution that should increase with growing magnetic field strength from central to peripheral collisions. On the other hand, the ratio for the scaled $\delta$ correlator also increases from unity in very central collisions to be above unity in more peripheral collisions, while a nonzero CME contribution would have decreased this ratio to be below unity. This apparent ``inconsistency'' between the $\gamma$ and $\delta$ trends requires a closer scrutiny of the background behavior.

To resolve this issue, we have used our $(20\sim 50)\%$ simulation events in the pure background case  to verify the assumption about the $1/N_{ch}$ background scaling. It turns out that the scaling is not exact and a non-negligible additional dependence on the average transverse momentum $\langle p_T \rangle$ can be identified, especially in the 
$\delta$ correlator. The physical origin of this effect is due to the initial fluctuations that could induce a spread of radial flow ``push" and thus a spread of $\langle p_T \rangle$ even for events with similar multiplicity. Stronger radial flow (i.e. larger $\langle p_T \rangle$) would lead to a stronger angular collimation of correlated charged hadron pairs, and thus to the enhancement of background correlations~\cite{Pratt:2010gy}. 

To demonstrate the impact of this effect on the $\delta$ and $\gamma$ correlators, we bin the $(20\sim 50)\%$  simulation events based on $\langle p_T \rangle$ and compute the corresponding correlators in each bin. The results, plotted in Fig.~\ref{fig_3}, clearly show a linear increase of $N_{ch} \times \Delta \delta$ with $\langle p_T \rangle$.  The  $N_{ch} \times \Delta \gamma / v_2$, on the other hand, appears to be relatively insensitive to the $\langle p_T \rangle$. We also note that hydrodynamic simulations performed in ~\cite{Nijs:2021kvn} and in our calculations demonstrate that the RuRu events have a larger $\langle p_T \rangle$ than ZrZr events in the same centrality class. 

Our findings suggest that while unity is a suitable baseline ratio of the $N_{ch} \times \Delta \gamma / v_2$ correlator, flow-induced corrections need to be taken into account for the baseline ratio of $N_{ch} \times \Delta \delta$. Since RuRu system has a larger multiplicity than ZrZr system in the same centrality class, the scaled $\delta$ correlator would  have a relative enhancement in the RuRu system due to a slightly larger radial flow ``push". To quantify this correction, we have  evaluated the baseline ratio from pure background case to be $1.033$ by using the $(20\sim 50)\%$ AVFD simulation events for the isobar pairs. In short, our analysis of the baseline ratios can be summarized as $R_\gamma^{baseline}=1$ and $R_\delta^{baseline}=1.033$ for  $(20\sim 50)\%$ collisions, shown as blue lines in Fig.~\ref{fig_2} (lower panels). Comparing with the corresponding STAR data, $R_\gamma^{STAR}= 1.011 > R_\gamma^{baseline}$ and $R_\delta^{STAR}= 1.028 < R_\delta^{baseline} $, shown as red bands in Fig.~\ref{fig_2} (lower panels), we conclude that both the scaled $\gamma$ and $\delta$ correlators are consistent with a nonzero CME contribution.

{\em Extracting CME signal fraction.---} Given the indication of a nonzero CME signal from our analysis above, we now make an attempt to extract the CME signal fraction  from both the $\gamma$ and $\delta$ correlators based on the available information for $(20\sim 50)\%$ centrality from the STAR data as well as from the simulation results.

Let us first examine the $\Delta {\bar{\gamma}}$ correlator. Assuming that the CME signal fraction is $f_s$, we split the correlator measured in  RuRu collisions into a signal and background contributions as follows:
\begin{eqnarray}
&&\Delta \bar{\gamma}^{Ru}_{s} = f_s  \Delta\bar{\gamma}^{Ru},  \\
&& \Delta \bar{\gamma}^{Ru}_{b} = (1-f_s) \Delta\bar{\gamma}^{Ru},
\end{eqnarray} 
where the subscripts ``s'' and ``b'' denote the signal and background components, respectively. Since the ZrZr collisions are expected to possess a weaker magnetic field, and thus relatively smaller signal and larger background, we then re-scale these quantities from RuRu to ZrZr collisions as follows: 
\begin{eqnarray}
&& \Delta\bar{\gamma}^{Zr}_{s} = (1-  \lambda_s)   \Delta\bar{\gamma}^{Ru}_{s} =   (1-  \lambda_s) f_s \Delta\bar{\gamma}^{Ru},  \\
&& \Delta\bar{\gamma}^{Zr}_{b} = (1 + \lambda_b) \Delta\bar{\gamma}^{Ru}_{b} =  (1 + \lambda_b)  (1-f_s) \Delta\bar{\gamma}^{Ru} .
\end{eqnarray}  
Therefore, the total $\Delta\bar{\gamma}$ correlator for ZrZr is:
\begin{eqnarray}
 \Delta\bar{\gamma}^{Zr}  &=&  \Delta\bar{\gamma}^{Ru}  \times \left[  (1-  \lambda_s) f_s  + (1 + \lambda_b)  (1-f_s) \right ] \nonumber  \\
 &=&   \Delta\bar{\gamma}^{Ru}  \left[ 1+ \lambda_b - (\lambda_s+\lambda_b) f_s \right ],
\end{eqnarray}
which means the $\Delta\bar{\gamma}$-ratio  between the isobars is 
\begin{eqnarray}
R  =  \frac{\Delta\bar{\gamma}^{Ru}}{\Delta\bar{\gamma}^{Zr} } = \frac{1}{1+ \lambda_b - (\lambda_s+\lambda_b) f_s} \, .
\end{eqnarray}
Let us rewrite the ratio as $R\equiv \frac{1}{1+\lambda_R}$ (or equivalently $\lambda_R=1/R-1$) and then we can express the $f_s$ as
\begin{eqnarray}
f_s = \frac{\lambda_b - \lambda_R}{\lambda_s + \lambda_b}.
\end{eqnarray} 
Under the na\"ive assumption of identical backgrounds in RuRu and ZrZr systems, as used in the STAR predefined criteria, the pure background case would correspond to $\lambda_b=\lambda_R=f_s=0$, while a nonzero signal would correspond to $\lambda_b=0$, $\lambda_R<0$ and $f_s>0$. This assumption is however invalided by the data, as already discussed above. 

\begin{figure}[!hbt]
	\begin{center}
		\includegraphics[width=3.2in]{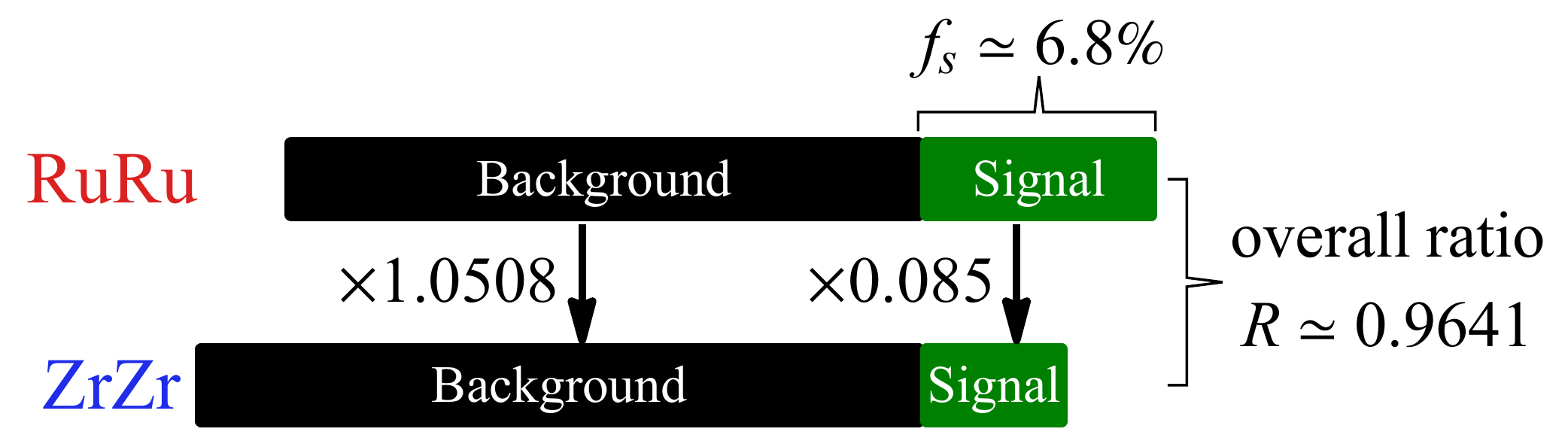}
		\caption{ An illustration of the comparison between isobar systems for the 
		measured $\bar{\gamma}$ correlators, with signal and background components indicated as black and green bars. The lengths of the bars are not plotted in exact proportion and the signal parts are graphically magnified for visibility. See text for details.}
		\vspace{-0.5cm}
		\label{fig_4}
	\end{center}
\end{figure}

The estimates based on experimental data and simulations suggest instead the following values: (a)  $R\simeq 0.9641 \pm 0.0037$ or $\lambda_R\simeq +(0.0372 \pm 0.0040)$ directly from measurements; 
(b) $\lambda_b$, dominated by the multiplicity difference, is estimated according to the ratio of $\langle N^{-1} \rangle$ and $\lambda_b\simeq +0.0508$; 
(c) The isobar signal ratio is dictated by the magnetic fields, for which the ratio is determined from simulations to be $\lambda_s \simeq +(0.15 \pm 0.05)$~\cite{Shi:2018sah}
. See Supplemental Material~\ref{sup}~\cite{SuppMat} for details of how these values are obtained or estimated. Putting these inputs together, one arrives at the following estimate for the CME signal fraction in the measured $\Delta\bar{\gamma}$ correlator: 
\begin{eqnarray}
f_s \simeq + (0.068 \pm 0.026 ) = (6.8 \pm 2.6 ) \%.
\end{eqnarray}
 We note that this value is consistent with the isobar analysis results from the event-plane/spectator-plane contrast method~\cite{STAR:2021mii}. To make the outcome of this analysis more transparent, we illustrate it in Fig.~\ref{fig_4}.

{\em Summary.---}
To summarize, we have combined the insights from theoretical simulations with the analysis of STAR experimental data to understand the implications of isobar collisions for the chiral magnetic effect. First, we have shown that the measured  multiplicity difference between the RuRu and ZrZr systems, which plays a key role for establishing the background baseline, can be successfully described by simulations with suitable initial nuclear structure inputs. Furthermore, we have identified the radial flow ``push" as an important contributor to background correlations, in addition to the multiplicity and elliptic flow. Quantitatively accounting for these two effects on the backgrounds has allowed us a calibration of the appropriate baselines for both the scaled $\gamma$ and $\delta$ correlators. Compared with the experimental data, we conclude that  the correlation measurements could be consistent with a finite CME signal contribution, estimated at a level of about $(6.8\pm2.6)\%$ fraction as illustrated in Fig.~\ref{fig_4}. 
Such a fraction is obtained by assuming that the non-CME background of $\Delta\bar{\gamma}$ is inversely proportional to multiplicity. There is however non-flow effect~\cite{Feng:2021pgf} that could make nontrivial contributions to background ratio, the influence of which clearly deserves future investigations (--- see Supplemental Material~\ref{sup}~\cite{SuppMat} for more discussions). 

Let us discuss possible future measurements that can further help to establish or rule out the presence of CME signal in heavy ion collisions. Given that the radial flow push is found to bear impact on the  calibration of backgrounds, it would be very useful to measure and compare the average transverse momentum of charged particles between the isobar systems. Another useful approach for the isobar comparison is to apply identical event selection criteria for multiplicity, $v_2$ and $\langle p_T \rangle$ and then compare the subset of isobar events that are ensured to have identical background correlations~\cite{Shi:2019wzi,Shi:2018sah}. Both the participant-plane/spectator-plane contrast method~\cite{Xu:2017qfs,Voloshin:2018qsm} and the  event-shape engineering approach~\cite{Milton:2021wku} have the potential of maximizing the signal/background contrast, extracting the signal fraction and allowing a verification of expected signal scaling between isobar pairs. Beyond the $\gamma$ and $\delta$ correlators, it would be interesting to understand the background scaling and their implications for the interpretation of isobar comparison in e.g. the R-correlator~\cite{Lacey:2022baf} as well as the signed balance function~\cite{Tang:2019pbl}. Finally, it is of great importance to achieve a coherent understanding of both isobar and AuAu systems~\cite{Feng:2021oub}. Recent STAR measurements in AuAu collisions based on  the participant-plane/spectator-plane contrast method, while suffering from a limited statistics, do suggest a nonzero CME fraction of $(6.3\sim 14.7)\%$~\cite{STAR:2021pwb}. 
Future high precision measurements based on the anticipated large sample of AuAu events together with the ongoing post-blinding analysis of the extensive isobar data set will hopefully allow to improve the statistical significance of these results and allow to make a conclusive statement on the presence or absence of CME in heavy ion collisions. 

\section*{Acknowledgments}
We thank H. Caines, Y. Hu, H. Huang, R. Lacey, S. Pratt, A. Tang, P. Tribedy, S. Voloshin, F. Wang, G. Wang and Z. Xu for useful discussions and communications. 
 This work is partly supported by the U.S. Department of Energy, Office of Science, Office of Nuclear Physics, within the framework of the Beam Energy Scan Theory (BEST) Topical Collaboration. D.K. and S.S. also acknowledge support by the U.S. Department of Energy, Office of Science, Office of Nuclear Physics, grants Nos. DE-FG88ER40388 and DE-SC0012704. J.L. also acknowledges support by the NSF Grant No. PHY-2209183. 

\clearpage
\begin{appendix}
\section{Supplementary Materials}\label{sup}

In this Supplemental Material, we provide a number of technical details on the simulations and analysis that have been used to obtain the results reported in the main manuscript. 

\subsubsection{The EBE-AVFD framework}

The EBE-AVFD model~\cite{Shi:2017cpu,Jiang:2016wve,Shi:2019wzi,An:2021wof} is a comprehensive simulation framework that dynamically describes  the CME in heavy-ion collisions. This state-of-the-art tool has been developed over the past few years to offer a quantitative and realistic characterization of the CME signals as well as the relevant backgrounds. Accordingly,  EBE-AVFD  implements the dynamical CME transport for quark currents on top of the relativistically expanding viscous QGP fluid and properly models   major sources of background correlations such as local charge conservation (LCC) and resonance decays. This tool has been extensively used by experimentalists at both RHIC and LHC to study CME-related observables and help their analyses~\cite{Choudhury:2021jwd,Christakoglou:2021nhe}. 

More specifically, the EBE-AVFD framework starts with event-wise fluctuating initial conditions, and solves the evolution of chiral quark currents as linear perturbations in addition to the viscous bulk flow background provided by data-validated hydrodynamic simulation packages. The  LCC effect is incorporated in the freeze-out process, followed by the hadron cascade simulations. 
The fluctuating initial conditions for entropy density profiles are generated by the Monte-Carlo Glauber model. 
The initial axial charge density ($n_5$) is approximated in such a way that it is proportional to  the corresponding local entropy density with a constant ratio. This ratio parameter can be varied to sensitively control the strength of the CME transport. For example, one can set  $n_5/s$ to $0$, $0.1$ and $0.2$  in the simulations to represent scenarios of zero, modest and strong CME signals respectively. The initial electromagnetic field is computed according to the event-wise proton configuration in the Monte-Carlo Glauber initial conditions. 
The hydrodynamic evolution is solved through two components. The bulk-matter collective flow is described by the VISH2+1 simulation package~\cite{Shen:2014vra}, with the lattice equation of state \texttt{s95p-v1.2}, shear-viscosity $\eta/s=0.08$, and freeze-out temperature $T_\text{fo}= 160~$MeV. Such hydrodynamic    simulations of bulk flow have been extensively tested and validated with relevant experimental data. The dynamical CME transport is described by anomalous hydrodynamic equations for the quark chiral currents on top of the bulk flow background, where the magnetic-field-induced CME currents lead to a charge separation in the fireball. 
After the hydrodynamic stage, hadrons are locally produced in all fluid cells on the freeze-out hypersurface, using the Cooper-Frye procedure. 
In the freeze-out process, the LCC effect is implemented in the EBE-AVFD package by generalizing an earlier method  to mimic more realistically the impact of a finite charge-correlation length, by using a parameter $P_\text{LCC}$  to characterize the fraction of charged hadrons that are sampled in positive-negative pairs  with the rest of the hadrons sampled individually. Varying the parameter $P_\text{LCC}$ between 0 and 1 would tune the LCC contributions from none to its maximum. 
Finally, all the hadrons produced from the freeze-out hypersurface are further subject to hadron cascades through the UrQMD simulations, which  account for various hadron resonance decay processes and automatically include their contributions to the charge-dependent correlations. More details can be found in ~\cite{Shi:2017cpu,Jiang:2016wve,Shi:2019wzi,An:2021wof}.

For the present work, we've performed EBE-AVFD simulations with new inputs for the Ru and Zr nucleon distributions that are used in the Monte-Carlo Glauber initial conditions.  We tested several different choices based on recent literature~\cite{Zhang:2021kxj,STAR:2021mii,Xu:2021vpn} and were able to reasonably reproduce the observed multiplicity distributions and the ratio between isobar systems, as shown in  Fig.~1 of the main manuscript. After successful calibration with isobar multiplicity ratio,  these simulation events are  used for computations of observables in subsequent figures.

\subsubsection{The estimation of signal fraction}

Here we provide details on the various numbers and associated uncertainties that are involved in our estimation of the CME signal fraction. 

First of all, our analysis uses the official STAR data from their isobar publication~\cite{STAR:2021mii} which are now publicly available from ~\cite{hepdata.115993}. We list the relevant numbers in the following table. 

\begin{table*}[!hpbt]
    \centering
    \begin{tabular}{l|c c c c c|c}
    \hline\hline
       && $20-30\%$ & $30-40\%$ & $40-50\%$ & mean & ratio\\
    \hline
\multirow{2}{*}{$100\times\langle\frac{\Delta\gamma}{v_2}\rangle$}
    &RuRu&  
    0.3738(16) & 0.5314(22) & 0.7563(36) & 0.5538(15)
    &\multirow{2}{*}{0.9641(37)}\\
    &ZrZr&  
    0.3827(16) & 0.5464(22) & 0.7942(38) & 0.5744(16)\\
    \hline
\multirow{2}{*}{$\langle N\Delta\delta \rangle$}
    &RuRu&  
    0.21739(8) & 0.19034(8) & 0.16083(8) & 0.18952(4)
    &\multirow{2}{*}{1.02814(33)}\\
    &ZrZr& 
    0.21373(8) & 0.18499(8) & 0.15427(8) & 0.18433(4) \\
    \hline
\multirow{2}{*}{$\langle\frac{N\Delta\gamma}{v_2}\rangle$}
    &RuRu&   
    0.4704(20) & 0.4528(19) & 0.4228(20) & 0.4487(11)
    &\multirow{2}{*}{1.0117(36)}\\
    &ZrZr& 
    0.4683(19) & 0.4459(18) & 0.4163(20) & 0.4435(11)\\
    \hline
\multirow{2}{*}{$10\times\langle v_2\rangle$}
    &RuRu&   
    0.57195(5) & 0.63703(6) & 0.67319(8) & 0.62739(4)
    &\multirow{2}{*}{1.01513(8)}\\
    &ZrZr& 
    0.56515(4) & 0.62652(6) & 0.66244(8) & 0.61804(4) \\
    \hline
\multirow{2}{*}{$\langle N\rangle$}
    &RuRu&   
    125.84 & 85.22 & 55.91 & 88.99
    &\multirow{2}{*}{1.0413}\\
    &ZrZr& 
    122.35 & 81.62 & 52.41 & 85.46 \\
    \hline
\multirow{2}{*}{$\langle 1/N\rangle$}
    &RuRu&   
     & & & 0.0126034
    &\multirow{2}{*}{0.94916}\\
    &ZrZr& 
     & & & 0.0132784 \\
    \hline
    \hline
    \end{tabular}
    \caption{Centrality and system dependent scaled correlators. Numbers in the parentheses indicate the uncertainties of the last digit or the last two digits. Uncertainties for $\langle N\rangle$ are negligibly small. Data from~\cite{hepdata.115993}. $\langle1/N\rangle$ is computed from the distribution function of multiplicity $\langle1/N\rangle = \big[\sum_{N=\min}^{\max} N^{-1} P(N)\big]\big/\big[\sum_{N=\min}^{\max} P(N)\big]$.
    }
    \label{tab:correlators}
\end{table*}

We next show the details for estimating the signal fraction. For this purpose, three numbers need to be obtained: the CME-driven signal ratio between Ru and Zr, the non-CME pure background ratio between Ru and Zr, as well as the overall measured correlator ratio between Ru and Zr.

~\\
\emph{CME-driven signal ratio} --- 
The CME-driven signal difference in $\Delta\gamma$ is dominated by the magnetic field squared projected along the second-order event plane, $\langle B^2 \cos(2\Psi_B-2\Psi_{2}+\pi) \rangle$. Therefore the signal ratio parameter  $\lambda_s \equiv \Delta\bar{\gamma}_s^{\text{Ru}}/\Delta\bar{\gamma}_s^{\text{Zr}} - 1 = R_{B} / R_{v_2} - 1$, with $R_{v_2}$ being the ratio of elliptic flow and $R_B$ the ratio of projected magnetic field squared projected. The theoretical uncertainty of $\lambda_s$ is dominated by that of the proton density distributions within the colliding nuclei. According to~\cite{Shi:2018sah} and similar calculations, one can estimate that $R_{B} = 1.17 \pm 0.05$. The measured $v_2$ ratio can be obtained from experimental data, $R_{v_2} = 1.0151$ for which the uncertainty is negligibly small.  Combining these together, we obtain the CME-driven signal ratio parameter to be  $\lambda_s = 0.15 \pm 0.05$.

~\\
\emph{Non-CME background ratio} ---  
In this work, we take the working assumption that non-CME pure background contribution to the correlator  $\Delta\bar{\gamma}$ is proportional to inverse multiplicity $1/N$. With the STAR data for multiplicity distributions available from the HEP database~\cite{hepdata.115993}, we compute the mean inverse multiplicity as $\langle1/N\rangle \equiv \frac{\sum_{N=\min}^{\max} N^{-1} P(N)}{\sum_{N=\min}^{\max} P(N)}$ and find that  $\langle1/N\rangle_\text{Ru} = 0.01260$ and $\langle1/N\rangle_\text{Zr} = 0.01328$ for the $(20-50)\%$ cenrality class. Accordingly, the estimation of the background ratio parameter is given by $\lambda_b = 1- R_{\langle1/N\rangle} = 0.05084$.  
As a caveat, there could be non-flow correction to the multiplicity scaling of background ratio used above, as demonstrated by recent analysis in \cite{Feng:2021pgf}. If such correction is estimated at a level of about $\pm 0.01$ uncertainty for the $\lambda_b$, it would amount to about $100\%$ of uncertainty in the extracted $f_s$. Clearly, this is an important factor that needs to be carefully quantified with further theoretical modelings and experimental analysis.

~\\
\emph{Calculation of the correlator ratio} ---  
For the correlator ratio $R$ of the scaled correlator $\Delta \bar{\gamma}$, we can simply read off the  numbers listed in Table~\ref{tab:correlators} to obtain   $R = 0.9641 \pm 0.0037$. Similarly, the RuRu-to-ZrZr ratio of multiplicity scaled correlators $N\Delta\delta$ and $N\Delta\gamma/v_2$ are found to be $1.02814 \pm 0.00033$ and $N\Delta\delta$ and $1.0117 \pm 0.0036$, respectively. Here, we have used the uncorrected number of tracks ($N_\mathrm{trk}$) for calculating charged multiplicity, $v_2\{\text{TPC EP}\}$ for elliptic flow and $\Delta\gamma\{\text{TPC EP}\}$ for $\Delta\gamma$. The results for the $20-50\%$ class are obtained as the mean value from those of $20-30\%$, $30-40\%$, and $40-50\%$ centrality classes. \\

Substituting all these numbers together into the Eq.~(10) of the main manuscript, we obtain the final result for the extracted signal fraction: $f_s = (6.8\pm 2.6)\%$. 
It should be noted that the  estimation of correlator ratio  $R$ is based on 
the full TPC measurement by Group-1, which is between the maximum and minimum values obtained from other analysis groups. If one takes the maximum value of $R = 0.9733 \pm 0.0040$, from the 3PC-TPC measurement of Group-3, the fraction becomes $f_s = (11.7\pm3.6)\%$; whereas for the minimum value, from Group-2's SE-TPC measurement,  it gives $R = 0.9611 \pm 0.0070$ and results in $f_s = (5.2\pm4.0)\%$.

\end{appendix}

\bibliography{isobar.bib}

\begin{thebibliography}{76}%
\makeatletter
\providecommand \@ifxundefined [1]{%
 \@ifx{#1\undefined}
}%
\providecommand \@ifnum [1]{%
 \ifnum #1\expandafter \@firstoftwo
 \else \expandafter \@secondoftwo
 \fi
}%
\providecommand \@ifx [1]{%
 \ifx #1\expandafter \@firstoftwo
 \else \expandafter \@secondoftwo
 \fi
}%
\providecommand \natexlab [1]{#1}%
\providecommand \enquote  [1]{``#1''}%
\providecommand \bibnamefont  [1]{#1}%
\providecommand \bibfnamefont [1]{#1}%
\providecommand \citenamefont [1]{#1}%
\providecommand \href@noop [0]{\@secondoftwo}%
\providecommand \href [0]{\begingroup \@sanitize@url \@href}%
\providecommand \@href[1]{\@@startlink{#1}\@@href}%
\providecommand \@@href[1]{\endgroup#1\@@endlink}%
\providecommand \@sanitize@url [0]{\catcode `\\12\catcode `\$12\catcode
  `\&12\catcode `\#12\catcode `\^12\catcode `\_12\catcode `\%12\relax}%
\providecommand \@@startlink[1]{}%
\providecommand \@@endlink[0]{}%
\providecommand \url  [0]{\begingroup\@sanitize@url \@url }%
\providecommand \@url [1]{\endgroup\@href {#1}{\urlprefix }}%
\providecommand \urlprefix  [0]{URL }%
\providecommand \Eprint [0]{\href }%
\providecommand \doibase [0]{http://dx.doi.org/}%
\providecommand \selectlanguage [0]{\@gobble}%
\providecommand \bibinfo  [0]{\@secondoftwo}%
\providecommand \bibfield  [0]{\@secondoftwo}%
\providecommand \translation [1]{[#1]}%
\providecommand \BibitemOpen [0]{}%
\providecommand \bibitemStop [0]{}%
\providecommand \bibitemNoStop [0]{.\EOS\space}%
\providecommand \EOS [0]{\spacefactor3000\relax}%
\providecommand \BibitemShut  [1]{\csname bibitem#1\endcsname}%
\let\auto@bib@innerbib\@empty
\bibitem [{\citenamefont {Kharzeev}(2006)}]{Kharzeev:2004ey}%
  \BibitemOpen
  \bibfield  {author} {\bibinfo {author} {\bibfnamefont {Dmitri}\ \bibnamefont
  {Kharzeev}},\ }\bibfield  {title} {\enquote {\bibinfo {title} {{Parity
  violation in hot QCD: Why it can happen, and how to look for it}},}\ }\href
  {\doibase 10.1016/j.physletb.2005.11.075} {\bibfield  {journal} {\bibinfo
  {journal} {Phys. Lett. B}\ }\textbf {\bibinfo {volume} {633}},\ \bibinfo
  {pages} {260--264} (\bibinfo {year} {2006})},\ \Eprint
  {http://arxiv.org/abs/hep-ph/0406125} {arXiv:hep-ph/0406125} \BibitemShut
  {NoStop}%
\bibitem [{\citenamefont {Kharzeev}\ \emph {et~al.}(2008)\citenamefont
  {Kharzeev}, \citenamefont {McLerran},\ and\ \citenamefont
  {Warringa}}]{Kharzeev:2007jp}%
  \BibitemOpen
  \bibfield  {author} {\bibinfo {author} {\bibfnamefont {Dmitri~E.}\
  \bibnamefont {Kharzeev}}, \bibinfo {author} {\bibfnamefont {Larry~D.}\
  \bibnamefont {McLerran}}, \ and\ \bibinfo {author} {\bibfnamefont
  {Harmen~J.}\ \bibnamefont {Warringa}},\ }\bibfield  {title} {\enquote
  {\bibinfo {title} {{The Effects of topological charge change in heavy ion
  collisions: 'Event by event P and CP violation'}},}\ }\href {\doibase
  10.1016/j.nuclphysa.2008.02.298} {\bibfield  {journal} {\bibinfo  {journal}
  {Nucl. Phys. A}\ }\textbf {\bibinfo {volume} {803}},\ \bibinfo {pages}
  {227--253} (\bibinfo {year} {2008})},\ \Eprint
  {http://arxiv.org/abs/0711.0950} {arXiv:0711.0950 [hep-ph]} \BibitemShut
  {NoStop}%
\bibitem [{\citenamefont {Fukushima}\ \emph {et~al.}(2008)\citenamefont
  {Fukushima}, \citenamefont {Kharzeev},\ and\ \citenamefont
  {Warringa}}]{Fukushima:2008xe}%
  \BibitemOpen
  \bibfield  {author} {\bibinfo {author} {\bibfnamefont {Kenji}\ \bibnamefont
  {Fukushima}}, \bibinfo {author} {\bibfnamefont {Dmitri~E.}\ \bibnamefont
  {Kharzeev}}, \ and\ \bibinfo {author} {\bibfnamefont {Harmen~J.}\
  \bibnamefont {Warringa}},\ }\bibfield  {title} {\enquote {\bibinfo {title}
  {{The Chiral Magnetic Effect}},}\ }\href {\doibase
  10.1103/PhysRevD.78.074033} {\bibfield  {journal} {\bibinfo  {journal} {Phys.
  Rev. D}\ }\textbf {\bibinfo {volume} {78}},\ \bibinfo {pages} {074033}
  (\bibinfo {year} {2008})},\ \Eprint {http://arxiv.org/abs/0808.3382}
  {arXiv:0808.3382 [hep-ph]} \BibitemShut {NoStop}%
\bibitem [{\citenamefont {Kharzeev}\ \emph {et~al.}(1998)\citenamefont
  {Kharzeev}, \citenamefont {Pisarski},\ and\ \citenamefont
  {Tytgat}}]{Kharzeev:1998kz}%
  \BibitemOpen
  \bibfield  {author} {\bibinfo {author} {\bibfnamefont {Dmitri}\ \bibnamefont
  {Kharzeev}}, \bibinfo {author} {\bibfnamefont {R.~D.}\ \bibnamefont
  {Pisarski}}, \ and\ \bibinfo {author} {\bibfnamefont {Michel H.~G.}\
  \bibnamefont {Tytgat}},\ }\bibfield  {title} {\enquote {\bibinfo {title}
  {{Possibility of spontaneous parity violation in hot QCD}},}\ }\href
  {\doibase 10.1103/PhysRevLett.81.512} {\bibfield  {journal} {\bibinfo
  {journal} {Phys. Rev. Lett.}\ }\textbf {\bibinfo {volume} {81}},\ \bibinfo
  {pages} {512--515} (\bibinfo {year} {1998})},\ \Eprint
  {http://arxiv.org/abs/hep-ph/9804221} {arXiv:hep-ph/9804221} \BibitemShut
  {NoStop}%
\bibitem [{\citenamefont {Kharzeev}\ \emph {et~al.}(2002)\citenamefont
  {Kharzeev}, \citenamefont {Krasnitz},\ and\ \citenamefont
  {Venugopalan}}]{Kharzeev:2001ev}%
  \BibitemOpen
  \bibfield  {author} {\bibinfo {author} {\bibfnamefont {D.}~\bibnamefont
  {Kharzeev}}, \bibinfo {author} {\bibfnamefont {A.}~\bibnamefont {Krasnitz}},
  \ and\ \bibinfo {author} {\bibfnamefont {R.}~\bibnamefont {Venugopalan}},\
  }\bibfield  {title} {\enquote {\bibinfo {title} {{Anomalous chirality
  fluctuations in the initial stage of heavy ion collisions and parity odd
  bubbles}},}\ }\href {\doibase 10.1016/S0370-2693(02)02630-8} {\bibfield
  {journal} {\bibinfo  {journal} {Phys. Lett. B}\ }\textbf {\bibinfo {volume}
  {545}},\ \bibinfo {pages} {298--306} (\bibinfo {year} {2002})},\ \Eprint
  {http://arxiv.org/abs/hep-ph/0109253} {arXiv:hep-ph/0109253} \BibitemShut
  {NoStop}%
\bibitem [{\citenamefont {Kharzeev}\ and\ \citenamefont
  {Zhitnitsky}(2007)}]{Kharzeev:2007tn}%
  \BibitemOpen
  \bibfield  {author} {\bibinfo {author} {\bibfnamefont {D.}~\bibnamefont
  {Kharzeev}}\ and\ \bibinfo {author} {\bibfnamefont {A.}~\bibnamefont
  {Zhitnitsky}},\ }\bibfield  {title} {\enquote {\bibinfo {title} {{Charge
  separation induced by P-odd bubbles in QCD matter}},}\ }\href {\doibase
  10.1016/j.nuclphysa.2007.10.001} {\bibfield  {journal} {\bibinfo  {journal}
  {Nucl. Phys. A}\ }\textbf {\bibinfo {volume} {797}},\ \bibinfo {pages}
  {67--79} (\bibinfo {year} {2007})},\ \Eprint {http://arxiv.org/abs/0706.1026}
  {arXiv:0706.1026 [hep-ph]} \BibitemShut {NoStop}%
\bibitem [{\citenamefont {Son}\ and\ \citenamefont
  {Spivak}(2013)}]{Son:2012bg}%
  \BibitemOpen
  \bibfield  {author} {\bibinfo {author} {\bibfnamefont {D.~T.}\ \bibnamefont
  {Son}}\ and\ \bibinfo {author} {\bibfnamefont {B.~Z.}\ \bibnamefont
  {Spivak}},\ }\bibfield  {title} {\enquote {\bibinfo {title} {{Chiral Anomaly
  and Classical Negative Magnetoresistance of Weyl Metals}},}\ }\href {\doibase
  10.1103/PhysRevB.88.104412} {\bibfield  {journal} {\bibinfo  {journal} {Phys.
  Rev. B}\ }\textbf {\bibinfo {volume} {88}},\ \bibinfo {pages} {104412}
  (\bibinfo {year} {2013})},\ \Eprint {http://arxiv.org/abs/1206.1627}
  {arXiv:1206.1627 [cond-mat.mes-hall]} \BibitemShut {NoStop}%
\bibitem [{\citenamefont {Zyuzin}\ and\ \citenamefont
  {Burkov}(2012)}]{Zyuzin:2012tv}%
  \BibitemOpen
  \bibfield  {author} {\bibinfo {author} {\bibfnamefont {A.~A.}\ \bibnamefont
  {Zyuzin}}\ and\ \bibinfo {author} {\bibfnamefont {A.~A.}\ \bibnamefont
  {Burkov}},\ }\bibfield  {title} {\enquote {\bibinfo {title} {{Topological
  response in Weyl semimetals and the chiral anomaly}},}\ }\href {\doibase
  10.1103/PhysRevB.86.115133} {\bibfield  {journal} {\bibinfo  {journal} {Phys.
  Rev. B}\ }\textbf {\bibinfo {volume} {86}},\ \bibinfo {pages} {115133}
  (\bibinfo {year} {2012})},\ \Eprint {http://arxiv.org/abs/1206.1868}
  {arXiv:1206.1868 [cond-mat.mes-hall]} \BibitemShut {NoStop}%
\bibitem [{\citenamefont {Basar}\ \emph {et~al.}(2014)\citenamefont {Basar},
  \citenamefont {Kharzeev},\ and\ \citenamefont {Yee}}]{Basar:2013iaa}%
  \BibitemOpen
  \bibfield  {author} {\bibinfo {author} {\bibfnamefont {Gokce}\ \bibnamefont
  {Basar}}, \bibinfo {author} {\bibfnamefont {Dmitri~E.}\ \bibnamefont
  {Kharzeev}}, \ and\ \bibinfo {author} {\bibfnamefont {Ho-Ung}\ \bibnamefont
  {Yee}},\ }\bibfield  {title} {\enquote {\bibinfo {title} {{Triangle anomaly
  in Weyl semimetals}},}\ }\href {\doibase 10.1103/PhysRevB.89.035142}
  {\bibfield  {journal} {\bibinfo  {journal} {Phys. Rev. B}\ }\textbf {\bibinfo
  {volume} {89}},\ \bibinfo {pages} {035142} (\bibinfo {year} {2014})},\
  \Eprint {http://arxiv.org/abs/1305.6338} {arXiv:1305.6338 [hep-th]}
  \BibitemShut {NoStop}%
\bibitem [{\citenamefont {Li}\ \emph {et~al.}(2016)\citenamefont {Li},
  \citenamefont {Kharzeev}, \citenamefont {Zhang}, \citenamefont {Huang},
  \citenamefont {Pletikosic}, \citenamefont {Fedorov}, \citenamefont {Zhong},
  \citenamefont {Schneeloch}, \citenamefont {Gu},\ and\ \citenamefont
  {Valla}}]{Li:2014bha}%
  \BibitemOpen
  \bibfield  {author} {\bibinfo {author} {\bibfnamefont {Qiang}\ \bibnamefont
  {Li}}, \bibinfo {author} {\bibfnamefont {Dmitri~E.}\ \bibnamefont
  {Kharzeev}}, \bibinfo {author} {\bibfnamefont {Cheng}\ \bibnamefont {Zhang}},
  \bibinfo {author} {\bibfnamefont {Yuan}\ \bibnamefont {Huang}}, \bibinfo
  {author} {\bibfnamefont {I.}~\bibnamefont {Pletikosic}}, \bibinfo {author}
  {\bibfnamefont {A.~V.}\ \bibnamefont {Fedorov}}, \bibinfo {author}
  {\bibfnamefont {R.~D.}\ \bibnamefont {Zhong}}, \bibinfo {author}
  {\bibfnamefont {J.~A.}\ \bibnamefont {Schneeloch}}, \bibinfo {author}
  {\bibfnamefont {G.~D.}\ \bibnamefont {Gu}}, \ and\ \bibinfo {author}
  {\bibfnamefont {T.}~\bibnamefont {Valla}},\ }\bibfield  {title} {\enquote
  {\bibinfo {title} {{Observation of the chiral magnetic effect in ZrTe5}},}\
  }\href {\doibase 10.1038/nphys3648} {\bibfield  {journal} {\bibinfo
  {journal} {Nature Phys.}\ }\textbf {\bibinfo {volume} {12}},\ \bibinfo
  {pages} {550--554} (\bibinfo {year} {2016})},\ \Eprint
  {http://arxiv.org/abs/1412.6543} {arXiv:1412.6543 [cond-mat.str-el]}
  \BibitemShut {NoStop}%
\bibitem [{\citenamefont {Huang}\ \emph {et~al.}(2015)\citenamefont {Huang}
  \emph {et~al.}}]{Huang:2015eia}%
  \BibitemOpen
  \bibfield  {author} {\bibinfo {author} {\bibfnamefont {Xiaochun}\
  \bibnamefont {Huang}} \emph {et~al.},\ }\bibfield  {title} {\enquote
  {\bibinfo {title} {{Observation of the Chiral-Anomaly-Induced Negative
  Magnetoresistance in 3D Weyl Semimetal TaAs}},}\ }\href {\doibase
  10.1103/PhysRevX.5.031023} {\bibfield  {journal} {\bibinfo  {journal} {Phys.
  Rev. X}\ }\textbf {\bibinfo {volume} {5}},\ \bibinfo {pages} {031023}
  (\bibinfo {year} {2015})},\ \Eprint {http://arxiv.org/abs/1503.01304}
  {arXiv:1503.01304 [cond-mat.mtrl-sci]} \BibitemShut {NoStop}%
\bibitem [{\citenamefont {Grabowska}\ \emph {et~al.}(2015)\citenamefont
  {Grabowska}, \citenamefont {Kaplan},\ and\ \citenamefont
  {Reddy}}]{Grabowska:2014efa}%
  \BibitemOpen
  \bibfield  {author} {\bibinfo {author} {\bibfnamefont {Dorota}\ \bibnamefont
  {Grabowska}}, \bibinfo {author} {\bibfnamefont {David~B.}\ \bibnamefont
  {Kaplan}}, \ and\ \bibinfo {author} {\bibfnamefont {Sanjay}\ \bibnamefont
  {Reddy}},\ }\bibfield  {title} {\enquote {\bibinfo {title} {{Role of the
  electron mass in damping chiral plasma instability in Supernovae and neutron
  stars}},}\ }\href {\doibase 10.1103/PhysRevD.91.085035} {\bibfield  {journal}
  {\bibinfo  {journal} {Phys. Rev. D}\ }\textbf {\bibinfo {volume} {91}},\
  \bibinfo {pages} {085035} (\bibinfo {year} {2015})},\ \Eprint
  {http://arxiv.org/abs/1409.3602} {arXiv:1409.3602 [hep-ph]} \BibitemShut
  {NoStop}%
\bibitem [{\citenamefont {Masada}\ \emph {et~al.}(2018)\citenamefont {Masada},
  \citenamefont {Kotake}, \citenamefont {Takiwaki},\ and\ \citenamefont
  {Yamamoto}}]{Masada:2018swb}%
  \BibitemOpen
  \bibfield  {author} {\bibinfo {author} {\bibfnamefont {Youhei}\ \bibnamefont
  {Masada}}, \bibinfo {author} {\bibfnamefont {Kei}\ \bibnamefont {Kotake}},
  \bibinfo {author} {\bibfnamefont {Tomoya}\ \bibnamefont {Takiwaki}}, \ and\
  \bibinfo {author} {\bibfnamefont {Naoki}\ \bibnamefont {Yamamoto}},\
  }\bibfield  {title} {\enquote {\bibinfo {title} {{Chiral magnetohydrodynamic
  turbulence in core-collapse supernovae}},}\ }\href {\doibase
  10.1103/PhysRevD.98.083018} {\bibfield  {journal} {\bibinfo  {journal} {Phys.
  Rev. D}\ }\textbf {\bibinfo {volume} {98}},\ \bibinfo {pages} {083018}
  (\bibinfo {year} {2018})},\ \Eprint {http://arxiv.org/abs/1805.10419}
  {arXiv:1805.10419 [astro-ph.HE]} \BibitemShut {NoStop}%
\bibitem [{\citenamefont {Yamamoto}(2016)}]{Yamamoto:2015gzz}%
  \BibitemOpen
  \bibfield  {author} {\bibinfo {author} {\bibfnamefont {Naoki}\ \bibnamefont
  {Yamamoto}},\ }\bibfield  {title} {\enquote {\bibinfo {title} {{Chiral
  transport of neutrinos in supernovae: Neutrino-induced fluid helicity and
  helical plasma instability}},}\ }\href {\doibase 10.1103/PhysRevD.93.065017}
  {\bibfield  {journal} {\bibinfo  {journal} {Phys. Rev. D}\ }\textbf {\bibinfo
  {volume} {93}},\ \bibinfo {pages} {065017} (\bibinfo {year} {2016})},\
  \Eprint {http://arxiv.org/abs/1511.00933} {arXiv:1511.00933 [astro-ph.HE]}
  \BibitemShut {NoStop}%
\bibitem [{\citenamefont {Tashiro}\ \emph {et~al.}(2012)\citenamefont
  {Tashiro}, \citenamefont {Vachaspati},\ and\ \citenamefont
  {Vilenkin}}]{Tashiro:2012mf}%
  \BibitemOpen
  \bibfield  {author} {\bibinfo {author} {\bibfnamefont {Hiroyuki}\
  \bibnamefont {Tashiro}}, \bibinfo {author} {\bibfnamefont {Tanmay}\
  \bibnamefont {Vachaspati}}, \ and\ \bibinfo {author} {\bibfnamefont
  {Alexander}\ \bibnamefont {Vilenkin}},\ }\bibfield  {title} {\enquote
  {\bibinfo {title} {{Chiral Effects and Cosmic Magnetic Fields}},}\ }\href
  {\doibase 10.1103/PhysRevD.86.105033} {\bibfield  {journal} {\bibinfo
  {journal} {Phys. Rev. D}\ }\textbf {\bibinfo {volume} {86}},\ \bibinfo
  {pages} {105033} (\bibinfo {year} {2012})},\ \Eprint
  {http://arxiv.org/abs/1206.5549} {arXiv:1206.5549 [astro-ph.CO]} \BibitemShut
  {NoStop}%
\bibitem [{\citenamefont {Vilenkin}\ and\ \citenamefont
  {Leahy}(1982)}]{Vilenkin:1982pn}%
  \BibitemOpen
  \bibfield  {author} {\bibinfo {author} {\bibfnamefont {A.}~\bibnamefont
  {Vilenkin}}\ and\ \bibinfo {author} {\bibfnamefont {D.~A.}\ \bibnamefont
  {Leahy}},\ }\bibfield  {title} {\enquote {\bibinfo {title} {{PARITY
  NONCONSERVATION AND THE ORIGIN OF COSMIC MAGNETIC FIELDS}},}\ }\href
  {\doibase 10.1086/159706} {\bibfield  {journal} {\bibinfo  {journal}
  {Astrophys. J.}\ }\textbf {\bibinfo {volume} {254}},\ \bibinfo {pages}
  {77--81} (\bibinfo {year} {1982})}\BibitemShut {NoStop}%
\bibitem [{\citenamefont {Vilenkin}(1980)}]{Vilenkin:1980fu}%
  \BibitemOpen
  \bibfield  {author} {\bibinfo {author} {\bibfnamefont {A.}~\bibnamefont
  {Vilenkin}},\ }\bibfield  {title} {\enquote {\bibinfo {title} {{EQUILIBRIUM
  PARITY VIOLATING CURRENT IN A MAGNETIC FIELD}},}\ }\href {\doibase
  10.1103/PhysRevD.22.3080} {\bibfield  {journal} {\bibinfo  {journal} {Phys.
  Rev. D}\ }\textbf {\bibinfo {volume} {22}},\ \bibinfo {pages} {3080--3084}
  (\bibinfo {year} {1980})}\BibitemShut {NoStop}%
\bibitem [{\citenamefont {Akamatsu}\ and\ \citenamefont
  {Yamamoto}(2013)}]{Akamatsu:2013pjd}%
  \BibitemOpen
  \bibfield  {author} {\bibinfo {author} {\bibfnamefont {Yukinao}\ \bibnamefont
  {Akamatsu}}\ and\ \bibinfo {author} {\bibfnamefont {Naoki}\ \bibnamefont
  {Yamamoto}},\ }\bibfield  {title} {\enquote {\bibinfo {title} {{Chiral Plasma
  Instabilities}},}\ }\href {\doibase 10.1103/PhysRevLett.111.052002}
  {\bibfield  {journal} {\bibinfo  {journal} {Phys. Rev. Lett.}\ }\textbf
  {\bibinfo {volume} {111}},\ \bibinfo {pages} {052002} (\bibinfo {year}
  {2013})},\ \Eprint {http://arxiv.org/abs/1302.2125} {arXiv:1302.2125
  [nucl-th]} \BibitemShut {NoStop}%
\bibitem [{\citenamefont {Hirono}\ \emph {et~al.}(2016)\citenamefont {Hirono},
  \citenamefont {Kharzeev},\ and\ \citenamefont {Yin}}]{Hirono:2016jps}%
  \BibitemOpen
  \bibfield  {author} {\bibinfo {author} {\bibfnamefont {Yuji}\ \bibnamefont
  {Hirono}}, \bibinfo {author} {\bibfnamefont {Dmitri~E.}\ \bibnamefont
  {Kharzeev}}, \ and\ \bibinfo {author} {\bibfnamefont {Yi}~\bibnamefont
  {Yin}},\ }\bibfield  {title} {\enquote {\bibinfo {title} {{Quantized chiral
  magnetic current from reconnections of magnetic flux}},}\ }\href {\doibase
  10.1103/PhysRevLett.117.172301} {\bibfield  {journal} {\bibinfo  {journal}
  {Phys. Rev. Lett.}\ }\textbf {\bibinfo {volume} {117}},\ \bibinfo {pages}
  {172301} (\bibinfo {year} {2016})},\ \Eprint
  {http://arxiv.org/abs/1606.09611} {arXiv:1606.09611 [hep-ph]} \BibitemShut
  {NoStop}%
\bibitem [{\citenamefont {Gorbar}\ \emph {et~al.}(2016)\citenamefont {Gorbar},
  \citenamefont {Shovkovy}, \citenamefont {Vilchinskii}, \citenamefont
  {Rudenok}, \citenamefont {Boyarsky},\ and\ \citenamefont
  {Ruchayskiy}}]{Gorbar:2016qfh}%
  \BibitemOpen
  \bibfield  {author} {\bibinfo {author} {\bibfnamefont {E.~V.}\ \bibnamefont
  {Gorbar}}, \bibinfo {author} {\bibfnamefont {I.~A.}\ \bibnamefont
  {Shovkovy}}, \bibinfo {author} {\bibfnamefont {S.}~\bibnamefont
  {Vilchinskii}}, \bibinfo {author} {\bibfnamefont {I.}~\bibnamefont
  {Rudenok}}, \bibinfo {author} {\bibfnamefont {A.}~\bibnamefont {Boyarsky}}, \
  and\ \bibinfo {author} {\bibfnamefont {O.}~\bibnamefont {Ruchayskiy}},\
  }\bibfield  {title} {\enquote {\bibinfo {title} {{Anomalous Maxwell equations
  for inhomogeneous chiral plasma}},}\ }\href {\doibase
  10.1103/PhysRevD.93.105028} {\bibfield  {journal} {\bibinfo  {journal} {Phys.
  Rev. D}\ }\textbf {\bibinfo {volume} {93}},\ \bibinfo {pages} {105028}
  (\bibinfo {year} {2016})},\ \Eprint {http://arxiv.org/abs/1603.03442}
  {arXiv:1603.03442 [hep-th]} \BibitemShut {NoStop}%
\bibitem [{\citenamefont {Kharzeev}\ and\ \citenamefont
  {Li}(2019)}]{Kharzeev:2019ceh}%
  \BibitemOpen
  \bibfield  {author} {\bibinfo {author} {\bibfnamefont {Dmitri~E.}\
  \bibnamefont {Kharzeev}}\ and\ \bibinfo {author} {\bibfnamefont {Qiang}\
  \bibnamefont {Li}},\ }\bibfield  {title} {\enquote {\bibinfo {title} {{The
  Chiral Qubit: quantum computing with chiral anomaly}},}\ }\href@noop {} {\
  (\bibinfo {year} {2019})},\ \Eprint {http://arxiv.org/abs/1903.07133}
  {arXiv:1903.07133 [quant-ph]} \BibitemShut {NoStop}%
\bibitem [{\citenamefont {Shevchenko}(2013)}]{Shevchenko:2012nv}%
  \BibitemOpen
  \bibfield  {author} {\bibinfo {author} {\bibfnamefont {V.}~\bibnamefont
  {Shevchenko}},\ }\bibfield  {title} {\enquote {\bibinfo {title} {{Quantum
  measurements and chiral magnetic effect}},}\ }\href {\doibase
  10.1016/j.nuclphysb.2013.01.004} {\bibfield  {journal} {\bibinfo  {journal}
  {Nucl. Phys. B}\ }\textbf {\bibinfo {volume} {870}},\ \bibinfo {pages}
  {1--15} (\bibinfo {year} {2013})},\ \Eprint {http://arxiv.org/abs/1208.0777}
  {arXiv:1208.0777 [hep-th]} \BibitemShut {NoStop}%
\bibitem [{\citenamefont {Kharzeev}\ and\ \citenamefont
  {Liao}(2021)}]{Kharzeev:2020jxw}%
  \BibitemOpen
  \bibfield  {author} {\bibinfo {author} {\bibfnamefont {Dmitri~E.}\
  \bibnamefont {Kharzeev}}\ and\ \bibinfo {author} {\bibfnamefont {Jinfeng}\
  \bibnamefont {Liao}},\ }\bibfield  {title} {\enquote {\bibinfo {title}
  {{Chiral magnetic effect reveals the topology of gauge fields in heavy-ion
  collisions}},}\ }\href {\doibase 10.1038/s42254-020-00254-6} {\bibfield
  {journal} {\bibinfo  {journal} {Nature Rev. Phys.}\ }\textbf {\bibinfo
  {volume} {3}},\ \bibinfo {pages} {55--63} (\bibinfo {year} {2021})},\ \Eprint
  {http://arxiv.org/abs/2102.06623} {arXiv:2102.06623 [hep-ph]} \BibitemShut
  {NoStop}%
\bibitem [{\citenamefont {Kharzeev}\ \emph {et~al.}(2016)\citenamefont
  {Kharzeev}, \citenamefont {Liao}, \citenamefont {Voloshin},\ and\
  \citenamefont {Wang}}]{Kharzeev:2015znc}%
  \BibitemOpen
  \bibfield  {author} {\bibinfo {author} {\bibfnamefont {D.~E.}\ \bibnamefont
  {Kharzeev}}, \bibinfo {author} {\bibfnamefont {J.}~\bibnamefont {Liao}},
  \bibinfo {author} {\bibfnamefont {S.~A.}\ \bibnamefont {Voloshin}}, \ and\
  \bibinfo {author} {\bibfnamefont {G.}~\bibnamefont {Wang}},\ }\bibfield
  {title} {\enquote {\bibinfo {title} {{Chiral magnetic and vortical effects in
  high-energy nuclear collisions\textemdash{}A status report}},}\ }\href
  {\doibase 10.1016/j.ppnp.2016.01.001} {\bibfield  {journal} {\bibinfo
  {journal} {Prog. Part. Nucl. Phys.}\ }\textbf {\bibinfo {volume} {88}},\
  \bibinfo {pages} {1--28} (\bibinfo {year} {2016})},\ \Eprint
  {http://arxiv.org/abs/1511.04050} {arXiv:1511.04050 [hep-ph]} \BibitemShut
  {NoStop}%
\bibitem [{\citenamefont {Kharzeev}(2014)}]{Kharzeev:2013ffa}%
  \BibitemOpen
  \bibfield  {author} {\bibinfo {author} {\bibfnamefont {Dmitri~E.}\
  \bibnamefont {Kharzeev}},\ }\bibfield  {title} {\enquote {\bibinfo {title}
  {{The Chiral Magnetic Effect and Anomaly-Induced Transport}},}\ }\href
  {\doibase 10.1016/j.ppnp.2014.01.002} {\bibfield  {journal} {\bibinfo
  {journal} {Prog. Part. Nucl. Phys.}\ }\textbf {\bibinfo {volume} {75}},\
  \bibinfo {pages} {133--151} (\bibinfo {year} {2014})},\ \Eprint
  {http://arxiv.org/abs/1312.3348} {arXiv:1312.3348 [hep-ph]} \BibitemShut
  {NoStop}%
\bibitem [{\citenamefont {Fukushima}(2019)}]{Fukushima:2018grm}%
  \BibitemOpen
  \bibfield  {author} {\bibinfo {author} {\bibfnamefont {Kenji}\ \bibnamefont
  {Fukushima}},\ }\bibfield  {title} {\enquote {\bibinfo {title} {{Extreme
  matter in electromagnetic fields and rotation}},}\ }\href {\doibase
  10.1016/j.ppnp.2019.04.001} {\bibfield  {journal} {\bibinfo  {journal} {Prog.
  Part. Nucl. Phys.}\ }\textbf {\bibinfo {volume} {107}},\ \bibinfo {pages}
  {167--199} (\bibinfo {year} {2019})},\ \Eprint
  {http://arxiv.org/abs/1812.08886} {arXiv:1812.08886 [hep-ph]} \BibitemShut
  {NoStop}%
\bibitem [{\citenamefont {Hattori}\ and\ \citenamefont
  {Huang}(2017)}]{Hattori:2016emy}%
  \BibitemOpen
  \bibfield  {author} {\bibinfo {author} {\bibfnamefont {Koichi}\ \bibnamefont
  {Hattori}}\ and\ \bibinfo {author} {\bibfnamefont {Xu-Guang}\ \bibnamefont
  {Huang}},\ }\bibfield  {title} {\enquote {\bibinfo {title} {{Novel quantum
  phenomena induced by strong magnetic fields in heavy-ion collisions}},}\
  }\href {\doibase 10.1007/s41365-016-0178-3} {\bibfield  {journal} {\bibinfo
  {journal} {Nucl. Sci. Tech.}\ }\textbf {\bibinfo {volume} {28}},\ \bibinfo
  {pages} {26} (\bibinfo {year} {2017})},\ \Eprint
  {http://arxiv.org/abs/1609.00747} {arXiv:1609.00747 [nucl-th]} \BibitemShut
  {NoStop}%
\bibitem [{\citenamefont {Gao}\ \emph {et~al.}(2020)\citenamefont {Gao},
  \citenamefont {Ma}, \citenamefont {Pu},\ and\ \citenamefont
  {Wang}}]{Gao:2020vbh}%
  \BibitemOpen
  \bibfield  {author} {\bibinfo {author} {\bibfnamefont {Jian-Hua}\
  \bibnamefont {Gao}}, \bibinfo {author} {\bibfnamefont {Guo-Liang}\
  \bibnamefont {Ma}}, \bibinfo {author} {\bibfnamefont {Shi}\ \bibnamefont
  {Pu}}, \ and\ \bibinfo {author} {\bibfnamefont {Qun}\ \bibnamefont {Wang}},\
  }\bibfield  {title} {\enquote {\bibinfo {title} {{Recent developments in
  chiral and spin polarization effects in heavy-ion collisions}},}\ }\href
  {\doibase 10.1007/s41365-020-00801-x} {\bibfield  {journal} {\bibinfo
  {journal} {Nucl. Sci. Tech.}\ }\textbf {\bibinfo {volume} {31}},\ \bibinfo
  {pages} {90} (\bibinfo {year} {2020})},\ \Eprint
  {http://arxiv.org/abs/2005.10432} {arXiv:2005.10432 [hep-ph]} \BibitemShut
  {NoStop}%
\bibitem [{\citenamefont {Burkov}(2015)}]{Burkov:2015hba}%
  \BibitemOpen
  \bibfield  {author} {\bibinfo {author} {\bibfnamefont {A.~A.}\ \bibnamefont
  {Burkov}},\ }\bibfield  {title} {\enquote {\bibinfo {title} {{Chiral anomaly
  and transport in Weyl metals}},}\ }\href {\doibase
  10.1088/0953-8984/27/11/113201} {\bibfield  {journal} {\bibinfo  {journal}
  {J. Phys. Condens. Matter}\ }\textbf {\bibinfo {volume} {27}},\ \bibinfo
  {pages} {113201} (\bibinfo {year} {2015})},\ \Eprint
  {http://arxiv.org/abs/1502.07609} {arXiv:1502.07609 [cond-mat.mes-hall]}
  \BibitemShut {NoStop}%
\bibitem [{\citenamefont {Armitage}\ \emph {et~al.}(2018)\citenamefont
  {Armitage}, \citenamefont {Mele},\ and\ \citenamefont
  {Vishwanath}}]{Armitage:2017cjs}%
  \BibitemOpen
  \bibfield  {author} {\bibinfo {author} {\bibfnamefont {N.~P.}\ \bibnamefont
  {Armitage}}, \bibinfo {author} {\bibfnamefont {E.~J.}\ \bibnamefont {Mele}},
  \ and\ \bibinfo {author} {\bibfnamefont {Ashvin}\ \bibnamefont
  {Vishwanath}},\ }\bibfield  {title} {\enquote {\bibinfo {title} {{Weyl and
  Dirac Semimetals in Three Dimensional Solids}},}\ }\href {\doibase
  10.1103/RevModPhys.90.015001} {\bibfield  {journal} {\bibinfo  {journal}
  {Rev. Mod. Phys.}\ }\textbf {\bibinfo {volume} {90}},\ \bibinfo {pages}
  {015001} (\bibinfo {year} {2018})},\ \Eprint
  {http://arxiv.org/abs/1705.01111} {arXiv:1705.01111 [cond-mat.str-el]}
  \BibitemShut {NoStop}%
\bibitem [{\citenamefont {Shovkovy}(2021)}]{Shovkovy:2021yyw}%
  \BibitemOpen
  \bibfield  {author} {\bibinfo {author} {\bibfnamefont {Igor~A.}\ \bibnamefont
  {Shovkovy}},\ }\bibfield  {title} {\enquote {\bibinfo {title} {{Anomalous
  plasma: chiral magnetic effect and all that}},}\ }\href@noop {} {\  (\bibinfo
  {year} {2021})},\ \Eprint {http://arxiv.org/abs/2111.11416} {arXiv:2111.11416
  [nucl-th]} \BibitemShut {NoStop}%
\bibitem [{\citenamefont {Voloshin}(2004)}]{Voloshin:2004vk}%
  \BibitemOpen
  \bibfield  {author} {\bibinfo {author} {\bibfnamefont {Sergei~A.}\
  \bibnamefont {Voloshin}},\ }\bibfield  {title} {\enquote {\bibinfo {title}
  {{Parity violation in hot QCD: How to detect it}},}\ }\href {\doibase
  10.1103/PhysRevC.70.057901} {\bibfield  {journal} {\bibinfo  {journal} {Phys.
  Rev. C}\ }\textbf {\bibinfo {volume} {70}},\ \bibinfo {pages} {057901}
  (\bibinfo {year} {2004})},\ \Eprint {http://arxiv.org/abs/hep-ph/0406311}
  {arXiv:hep-ph/0406311} \BibitemShut {NoStop}%
\bibitem [{\citenamefont {Abelev}\ \emph {et~al.}(2009)\citenamefont {Abelev}
  \emph {et~al.}}]{STAR:2009wot}%
  \BibitemOpen
  \bibfield  {author} {\bibinfo {author} {\bibfnamefont {B.~I.}\ \bibnamefont
  {Abelev}} \emph {et~al.} (\bibinfo {collaboration} {STAR}),\ }\bibfield
  {title} {\enquote {\bibinfo {title} {{Azimuthal Charged-Particle Correlations
  and Possible Local Strong Parity Violation}},}\ }\href {\doibase
  10.1103/PhysRevLett.103.251601} {\bibfield  {journal} {\bibinfo  {journal}
  {Phys. Rev. Lett.}\ }\textbf {\bibinfo {volume} {103}},\ \bibinfo {pages}
  {251601} (\bibinfo {year} {2009})},\ \Eprint {http://arxiv.org/abs/0909.1739}
  {arXiv:0909.1739 [nucl-ex]} \BibitemShut {NoStop}%
\bibitem [{\citenamefont {Abelev}\ \emph {et~al.}(2010)\citenamefont {Abelev}
  \emph {et~al.}}]{STAR:2009tro}%
  \BibitemOpen
  \bibfield  {author} {\bibinfo {author} {\bibfnamefont {B.~I.}\ \bibnamefont
  {Abelev}} \emph {et~al.} (\bibinfo {collaboration} {STAR}),\ }\bibfield
  {title} {\enquote {\bibinfo {title} {{Observation of charge-dependent
  azimuthal correlations and possible local strong parity violation in heavy
  ion collisions}},}\ }\href {\doibase 10.1103/PhysRevC.81.054908} {\bibfield
  {journal} {\bibinfo  {journal} {Phys. Rev. C}\ }\textbf {\bibinfo {volume}
  {81}},\ \bibinfo {pages} {054908} (\bibinfo {year} {2010})},\ \Eprint
  {http://arxiv.org/abs/0909.1717} {arXiv:0909.1717 [nucl-ex]} \BibitemShut
  {NoStop}%
\bibitem [{\citenamefont {Acharya}\ \emph {et~al.}(2020)\citenamefont {Acharya}
  \emph {et~al.}}]{ALICE:2020siw}%
  \BibitemOpen
  \bibfield  {author} {\bibinfo {author} {\bibfnamefont {Shreyasi}\
  \bibnamefont {Acharya}} \emph {et~al.} (\bibinfo {collaboration} {ALICE}),\
  }\bibfield  {title} {\enquote {\bibinfo {title} {{Constraining the Chiral
  Magnetic Effect with charge-dependent azimuthal correlations in Pb-Pb
  collisions at $ \sqrt{s_{\mathrm{NN}}} $ = 2.76 and 5.02 TeV}},}\ }\href
  {\doibase 10.1007/JHEP09(2020)160} {\bibfield  {journal} {\bibinfo  {journal}
  {JHEP}\ }\textbf {\bibinfo {volume} {09}},\ \bibinfo {pages} {160} (\bibinfo
  {year} {2020})},\ \Eprint {http://arxiv.org/abs/2005.14640} {arXiv:2005.14640
  [nucl-ex]} \BibitemShut {NoStop}%
\bibitem [{\citenamefont {Acharya}\ \emph {et~al.}(2018)\citenamefont {Acharya}
  \emph {et~al.}}]{ALICE:2017sss}%
  \BibitemOpen
  \bibfield  {author} {\bibinfo {author} {\bibfnamefont {Shreyasi}\
  \bibnamefont {Acharya}} \emph {et~al.} (\bibinfo {collaboration} {ALICE}),\
  }\bibfield  {title} {\enquote {\bibinfo {title} {{Constraining the magnitude
  of the Chiral Magnetic Effect with Event Shape Engineering in Pb-Pb
  collisions at $\sqrt{s_\mathrm{NN}}$ = 2.76 TeV}},}\ }\href {\doibase
  10.1016/j.physletb.2017.12.021} {\bibfield  {journal} {\bibinfo  {journal}
  {Phys. Lett. B}\ }\textbf {\bibinfo {volume} {777}},\ \bibinfo {pages}
  {151--162} (\bibinfo {year} {2018})},\ \Eprint
  {http://arxiv.org/abs/1709.04723} {arXiv:1709.04723 [nucl-ex]} \BibitemShut
  {NoStop}%
\bibitem [{\citenamefont {Khachatryan}\ \emph {et~al.}(2017)\citenamefont
  {Khachatryan} \emph {et~al.}}]{CMS:2016wfo}%
  \BibitemOpen
  \bibfield  {author} {\bibinfo {author} {\bibfnamefont {Vardan}\ \bibnamefont
  {Khachatryan}} \emph {et~al.} (\bibinfo {collaboration} {CMS}),\ }\bibfield
  {title} {\enquote {\bibinfo {title} {{Observation of charge-dependent
  azimuthal correlations in $p$-Pb collisions and its implication for the
  search for the chiral magnetic effect}},}\ }\href {\doibase
  10.1103/PhysRevLett.118.122301} {\bibfield  {journal} {\bibinfo  {journal}
  {Phys. Rev. Lett.}\ }\textbf {\bibinfo {volume} {118}},\ \bibinfo {pages}
  {122301} (\bibinfo {year} {2017})},\ \Eprint
  {http://arxiv.org/abs/1610.00263} {arXiv:1610.00263 [nucl-ex]} \BibitemShut
  {NoStop}%
\bibitem [{\citenamefont {Zhao}\ and\ \citenamefont
  {Wang}(2019)}]{Zhao:2019hta}%
  \BibitemOpen
  \bibfield  {author} {\bibinfo {author} {\bibfnamefont {Jie}\ \bibnamefont
  {Zhao}}\ and\ \bibinfo {author} {\bibfnamefont {Fuqiang}\ \bibnamefont
  {Wang}},\ }\bibfield  {title} {\enquote {\bibinfo {title} {{Experimental
  searches for the chiral magnetic effect in heavy-ion collisions}},}\ }\href
  {\doibase 10.1016/j.ppnp.2019.05.001} {\bibfield  {journal} {\bibinfo
  {journal} {Prog. Part. Nucl. Phys.}\ }\textbf {\bibinfo {volume} {107}},\
  \bibinfo {pages} {200--236} (\bibinfo {year} {2019})},\ \Eprint
  {http://arxiv.org/abs/1906.11413} {arXiv:1906.11413 [nucl-ex]} \BibitemShut
  {NoStop}%
\bibitem [{\citenamefont {Li}\ and\ \citenamefont {Wang}(2020)}]{Li:2020dwr}%
  \BibitemOpen
  \bibfield  {author} {\bibinfo {author} {\bibfnamefont {Wei}\ \bibnamefont
  {Li}}\ and\ \bibinfo {author} {\bibfnamefont {Gang}\ \bibnamefont {Wang}},\
  }\bibfield  {title} {\enquote {\bibinfo {title} {{Chiral Magnetic Effects in
  Nuclear Collisions}},}\ }\href {\doibase 10.1146/annurev-nucl-030220-065203}
  {\bibfield  {journal} {\bibinfo  {journal} {Ann. Rev. Nucl. Part. Sci.}\
  }\textbf {\bibinfo {volume} {70}},\ \bibinfo {pages} {293--321} (\bibinfo
  {year} {2020})},\ \Eprint {http://arxiv.org/abs/2002.10397} {arXiv:2002.10397
  [nucl-ex]} \BibitemShut {NoStop}%
\bibitem [{\citenamefont {Bzdak}\ \emph {et~al.}(2020)\citenamefont {Bzdak},
  \citenamefont {Esumi}, \citenamefont {Koch}, \citenamefont {Liao},
  \citenamefont {Stephanov},\ and\ \citenamefont {Xu}}]{Bzdak:2019pkr}%
  \BibitemOpen
  \bibfield  {author} {\bibinfo {author} {\bibfnamefont {Adam}\ \bibnamefont
  {Bzdak}}, \bibinfo {author} {\bibfnamefont {Shinichi}\ \bibnamefont {Esumi}},
  \bibinfo {author} {\bibfnamefont {Volker}\ \bibnamefont {Koch}}, \bibinfo
  {author} {\bibfnamefont {Jinfeng}\ \bibnamefont {Liao}}, \bibinfo {author}
  {\bibfnamefont {Mikhail}\ \bibnamefont {Stephanov}}, \ and\ \bibinfo {author}
  {\bibfnamefont {Nu}~\bibnamefont {Xu}},\ }\bibfield  {title} {\enquote
  {\bibinfo {title} {{Mapping the Phases of Quantum Chromodynamics with Beam
  Energy Scan}},}\ }\href {\doibase 10.1016/j.physrep.2020.01.005} {\bibfield
  {journal} {\bibinfo  {journal} {Phys. Rept.}\ }\textbf {\bibinfo {volume}
  {853}},\ \bibinfo {pages} {1--87} (\bibinfo {year} {2020})},\ \Eprint
  {http://arxiv.org/abs/1906.00936} {arXiv:1906.00936 [nucl-th]} \BibitemShut
  {NoStop}%
\bibitem [{\citenamefont {Wang}(2010)}]{Wang:2009kd}%
  \BibitemOpen
  \bibfield  {author} {\bibinfo {author} {\bibfnamefont {Fuqiang}\ \bibnamefont
  {Wang}},\ }\bibfield  {title} {\enquote {\bibinfo {title} {{Effects of
  Cluster Particle Correlations on Local Parity Violation Observables}},}\
  }\href {\doibase 10.1103/PhysRevC.81.064902} {\bibfield  {journal} {\bibinfo
  {journal} {Phys. Rev. C}\ }\textbf {\bibinfo {volume} {81}},\ \bibinfo
  {pages} {064902} (\bibinfo {year} {2010})},\ \Eprint
  {http://arxiv.org/abs/0911.1482} {arXiv:0911.1482 [nucl-ex]} \BibitemShut
  {NoStop}%
\bibitem [{\citenamefont {Pratt}(2010)}]{Pratt:2010gy}%
  \BibitemOpen
  \bibfield  {author} {\bibinfo {author} {\bibfnamefont {Scott}\ \bibnamefont
  {Pratt}},\ }\bibfield  {title} {\enquote {\bibinfo {title} {{Alternative
  Contributions to the Angular Correlations Observed at RHIC Associated with
  Parity Fluctuations}},}\ }\href@noop {} {\  (\bibinfo {year} {2010})},\
  \Eprint {http://arxiv.org/abs/1002.1758} {arXiv:1002.1758 [nucl-th]}
  \BibitemShut {NoStop}%
\bibitem [{\citenamefont {Bzdak}\ \emph {et~al.}(2010)\citenamefont {Bzdak},
  \citenamefont {Koch},\ and\ \citenamefont {Liao}}]{Bzdak:2009fc}%
  \BibitemOpen
  \bibfield  {author} {\bibinfo {author} {\bibfnamefont {Adam}\ \bibnamefont
  {Bzdak}}, \bibinfo {author} {\bibfnamefont {Volker}\ \bibnamefont {Koch}}, \
  and\ \bibinfo {author} {\bibfnamefont {Jinfeng}\ \bibnamefont {Liao}},\
  }\bibfield  {title} {\enquote {\bibinfo {title} {{Remarks on possible local
  parity violation in heavy ion collisions}},}\ }\href {\doibase
  10.1103/PhysRevC.81.031901} {\bibfield  {journal} {\bibinfo  {journal} {Phys.
  Rev. C}\ }\textbf {\bibinfo {volume} {81}},\ \bibinfo {pages} {031901}
  (\bibinfo {year} {2010})},\ \Eprint {http://arxiv.org/abs/0912.5050}
  {arXiv:0912.5050 [nucl-th]} \BibitemShut {NoStop}%
\bibitem [{\citenamefont {Bzdak}\ \emph {et~al.}(2011)\citenamefont {Bzdak},
  \citenamefont {Koch},\ and\ \citenamefont {Liao}}]{Bzdak:2010fd}%
  \BibitemOpen
  \bibfield  {author} {\bibinfo {author} {\bibfnamefont {Adam}\ \bibnamefont
  {Bzdak}}, \bibinfo {author} {\bibfnamefont {Volker}\ \bibnamefont {Koch}}, \
  and\ \bibinfo {author} {\bibfnamefont {Jinfeng}\ \bibnamefont {Liao}},\
  }\bibfield  {title} {\enquote {\bibinfo {title} {{Azimuthal correlations from
  transverse momentum conservation and possible local parity violation}},}\
  }\href {\doibase 10.1103/PhysRevC.83.014905} {\bibfield  {journal} {\bibinfo
  {journal} {Phys. Rev. C}\ }\textbf {\bibinfo {volume} {83}},\ \bibinfo
  {pages} {014905} (\bibinfo {year} {2011})},\ \Eprint
  {http://arxiv.org/abs/1008.4919} {arXiv:1008.4919 [nucl-th]} \BibitemShut
  {NoStop}%
\bibitem [{\citenamefont {Bzdak}\ \emph {et~al.}(2013)\citenamefont {Bzdak},
  \citenamefont {Koch},\ and\ \citenamefont {Liao}}]{Bzdak:2012ia}%
  \BibitemOpen
  \bibfield  {author} {\bibinfo {author} {\bibfnamefont {Adam}\ \bibnamefont
  {Bzdak}}, \bibinfo {author} {\bibfnamefont {Volker}\ \bibnamefont {Koch}}, \
  and\ \bibinfo {author} {\bibfnamefont {Jinfeng}\ \bibnamefont {Liao}},\
  }\bibfield  {title} {\enquote {\bibinfo {title} {{Charge-Dependent
  Correlations in Relativistic Heavy Ion Collisions and the Chiral Magnetic
  Effect}},}\ }\href {\doibase 10.1007/978-3-642-37305-3_19} {\bibfield
  {journal} {\bibinfo  {journal} {Lect. Notes Phys.}\ }\textbf {\bibinfo
  {volume} {871}},\ \bibinfo {pages} {503--536} (\bibinfo {year} {2013})},\
  \Eprint {http://arxiv.org/abs/1207.7327} {arXiv:1207.7327 [nucl-th]}
  \BibitemShut {NoStop}%
\bibitem [{\citenamefont {Choudhury}\ \emph {et~al.}(2022)\citenamefont
  {Choudhury} \emph {et~al.}}]{Choudhury:2021jwd}%
  \BibitemOpen
  \bibfield  {author} {\bibinfo {author} {\bibfnamefont {Subikash}\
  \bibnamefont {Choudhury}} \emph {et~al.},\ }\bibfield  {title} {\enquote
  {\bibinfo {title} {{Investigation of experimental observables in search of
  the chiral magnetic effect in heavy-ion collisions in the STAR experiment
  *}},}\ }\href {\doibase 10.1088/1674-1137/ac2a1f} {\bibfield  {journal}
  {\bibinfo  {journal} {Chin. Phys. C}\ }\textbf {\bibinfo {volume} {46}},\
  \bibinfo {pages} {014101} (\bibinfo {year} {2022})},\ \Eprint
  {http://arxiv.org/abs/2105.06044} {arXiv:2105.06044 [nucl-ex]} \BibitemShut
  {NoStop}%
\bibitem [{\citenamefont {Christakoglou}\ \emph {et~al.}(2021)\citenamefont
  {Christakoglou}, \citenamefont {Qiu},\ and\ \citenamefont
  {Staa}}]{Christakoglou:2021nhe}%
  \BibitemOpen
  \bibfield  {author} {\bibinfo {author} {\bibfnamefont {Panos}\ \bibnamefont
  {Christakoglou}}, \bibinfo {author} {\bibfnamefont {Shi}\ \bibnamefont
  {Qiu}}, \ and\ \bibinfo {author} {\bibfnamefont {Joey}\ \bibnamefont
  {Staa}},\ }\bibfield  {title} {\enquote {\bibinfo {title} {{Systematic study
  of the chiral magnetic effect with the AVFD model at LHC energies}},}\ }\href
  {\doibase 10.1140/epjc/s10052-021-09498-7} {\bibfield  {journal} {\bibinfo
  {journal} {Eur. Phys. J. C}\ }\textbf {\bibinfo {volume} {81}},\ \bibinfo
  {pages} {717} (\bibinfo {year} {2021})},\ \Eprint
  {http://arxiv.org/abs/2106.03537} {arXiv:2106.03537 [nucl-th]} \BibitemShut
  {NoStop}%
\bibitem [{\citenamefont {Voloshin}(2010)}]{Voloshin:2010ut}%
  \BibitemOpen
  \bibfield  {author} {\bibinfo {author} {\bibfnamefont {Sergei~A.}\
  \bibnamefont {Voloshin}},\ }\bibfield  {title} {\enquote {\bibinfo {title}
  {{Testing the Chiral Magnetic Effect with Central U+U collisions}},}\ }\href
  {\doibase 10.1103/PhysRevLett.105.172301} {\bibfield  {journal} {\bibinfo
  {journal} {Phys. Rev. Lett.}\ }\textbf {\bibinfo {volume} {105}},\ \bibinfo
  {pages} {172301} (\bibinfo {year} {2010})},\ \Eprint
  {http://arxiv.org/abs/1006.1020} {arXiv:1006.1020 [nucl-th]} \BibitemShut
  {NoStop}%
\bibitem [{\citenamefont {Koch}\ \emph {et~al.}(2017)\citenamefont {Koch},
  \citenamefont {Schlichting}, \citenamefont {Skokov}, \citenamefont
  {Sorensen}, \citenamefont {Thomas}, \citenamefont {Voloshin}, \citenamefont
  {Wang},\ and\ \citenamefont {Yee}}]{Skokov:2016yrj}%
  \BibitemOpen
  \bibfield  {author} {\bibinfo {author} {\bibfnamefont {Volker}\ \bibnamefont
  {Koch}}, \bibinfo {author} {\bibfnamefont {Soeren}\ \bibnamefont
  {Schlichting}}, \bibinfo {author} {\bibfnamefont {Vladimir}\ \bibnamefont
  {Skokov}}, \bibinfo {author} {\bibfnamefont {Paul}\ \bibnamefont {Sorensen}},
  \bibinfo {author} {\bibfnamefont {Jim}\ \bibnamefont {Thomas}}, \bibinfo
  {author} {\bibfnamefont {Sergei}\ \bibnamefont {Voloshin}}, \bibinfo {author}
  {\bibfnamefont {Gang}\ \bibnamefont {Wang}}, \ and\ \bibinfo {author}
  {\bibfnamefont {Ho-Ung}\ \bibnamefont {Yee}},\ }\bibfield  {title} {\enquote
  {\bibinfo {title} {{Status of the chiral magnetic effect and collisions of
  isobars}},}\ }\href {\doibase 10.1088/1674-1137/41/7/072001} {\bibfield
  {journal} {\bibinfo  {journal} {Chin. Phys. C}\ }\textbf {\bibinfo {volume}
  {41}},\ \bibinfo {pages} {072001} (\bibinfo {year} {2017})},\ \Eprint
  {http://arxiv.org/abs/1608.00982} {arXiv:1608.00982 [nucl-th]} \BibitemShut
  {NoStop}%
\bibitem [{\citenamefont {Adam}\ \emph {et~al.}(2021)\citenamefont {Adam} \emph
  {et~al.}}]{STAR:2019bjg}%
  \BibitemOpen
  \bibfield  {author} {\bibinfo {author} {\bibfnamefont {J.}~\bibnamefont
  {Adam}} \emph {et~al.} (\bibinfo {collaboration} {STAR}),\ }\bibfield
  {title} {\enquote {\bibinfo {title} {{Methods for a blind analysis of isobar
  data collected by the STAR collaboration}},}\ }\href {\doibase
  10.1007/s41365-021-00878-y} {\bibfield  {journal} {\bibinfo  {journal} {Nucl.
  Sci. Tech.}\ }\textbf {\bibinfo {volume} {32}},\ \bibinfo {pages} {48}
  (\bibinfo {year} {2021})},\ \Eprint {http://arxiv.org/abs/1911.00596}
  {arXiv:1911.00596 [nucl-ex]} \BibitemShut {NoStop}%
\bibitem [{\citenamefont {Abdallah}\ \emph
  {et~al.}(2022{\natexlab{a}})\citenamefont {Abdallah} \emph
  {et~al.}}]{STAR:2021mii}%
  \BibitemOpen
  \bibfield  {author} {\bibinfo {author} {\bibfnamefont {Mohamed}\ \bibnamefont
  {Abdallah}} \emph {et~al.} (\bibinfo {collaboration} {STAR}),\ }\bibfield
  {title} {\enquote {\bibinfo {title} {{Search for the chiral magnetic effect
  with isobar collisions at $\sqrt {s_{NN}}$=200 GeV by the STAR Collaboration
  at the BNL Relativistic Heavy Ion Collider}},}\ }\href {\doibase
  10.1103/PhysRevC.105.014901} {\bibfield  {journal} {\bibinfo  {journal}
  {Phys. Rev. C}\ }\textbf {\bibinfo {volume} {105}},\ \bibinfo {pages}
  {014901} (\bibinfo {year} {2022}{\natexlab{a}})},\ \Eprint
  {http://arxiv.org/abs/2109.00131} {arXiv:2109.00131 [nucl-ex]} \BibitemShut
  {NoStop}%
\bibitem [{\citenamefont {Shi}\ \emph {et~al.}(2020)\citenamefont {Shi},
  \citenamefont {Zhang}, \citenamefont {Hou},\ and\ \citenamefont
  {Liao}}]{Shi:2019wzi}%
  \BibitemOpen
  \bibfield  {author} {\bibinfo {author} {\bibfnamefont {Shuzhe}\ \bibnamefont
  {Shi}}, \bibinfo {author} {\bibfnamefont {Hui}\ \bibnamefont {Zhang}},
  \bibinfo {author} {\bibfnamefont {Defu}\ \bibnamefont {Hou}}, \ and\ \bibinfo
  {author} {\bibfnamefont {Jinfeng}\ \bibnamefont {Liao}},\ }\bibfield  {title}
  {\enquote {\bibinfo {title} {{Signatures of Chiral Magnetic Effect in the
  Collisions of Isobars}},}\ }\href {\doibase 10.1103/PhysRevLett.125.242301}
  {\bibfield  {journal} {\bibinfo  {journal} {Phys. Rev. Lett.}\ }\textbf
  {\bibinfo {volume} {125}},\ \bibinfo {pages} {242301} (\bibinfo {year}
  {2020})},\ \Eprint {http://arxiv.org/abs/1910.14010} {arXiv:1910.14010
  [nucl-th]} \BibitemShut {NoStop}%
\bibitem [{\citenamefont {Shi}\ \emph {et~al.}(2018)\citenamefont {Shi},
  \citenamefont {Jiang}, \citenamefont {Lilleskov},\ and\ \citenamefont
  {Liao}}]{Shi:2017cpu}%
  \BibitemOpen
  \bibfield  {author} {\bibinfo {author} {\bibfnamefont {Shuzhe}\ \bibnamefont
  {Shi}}, \bibinfo {author} {\bibfnamefont {Yin}\ \bibnamefont {Jiang}},
  \bibinfo {author} {\bibfnamefont {Elias}\ \bibnamefont {Lilleskov}}, \ and\
  \bibinfo {author} {\bibfnamefont {Jinfeng}\ \bibnamefont {Liao}},\ }\bibfield
   {title} {\enquote {\bibinfo {title} {{Anomalous Chiral Transport in Heavy
  Ion Collisions from Anomalous-Viscous Fluid Dynamics}},}\ }\href {\doibase
  10.1016/j.aop.2018.04.026} {\bibfield  {journal} {\bibinfo  {journal} {Annals
  Phys.}\ }\textbf {\bibinfo {volume} {394}},\ \bibinfo {pages} {50--72}
  (\bibinfo {year} {2018})},\ \Eprint {http://arxiv.org/abs/1711.02496}
  {arXiv:1711.02496 [nucl-th]} \BibitemShut {NoStop}%
\bibitem [{\citenamefont {Jiang}\ \emph {et~al.}(2018)\citenamefont {Jiang},
  \citenamefont {Shi}, \citenamefont {Yin},\ and\ \citenamefont
  {Liao}}]{Jiang:2016wve}%
  \BibitemOpen
  \bibfield  {author} {\bibinfo {author} {\bibfnamefont {Yin}\ \bibnamefont
  {Jiang}}, \bibinfo {author} {\bibfnamefont {Shuzhe}\ \bibnamefont {Shi}},
  \bibinfo {author} {\bibfnamefont {Yi}~\bibnamefont {Yin}}, \ and\ \bibinfo
  {author} {\bibfnamefont {Jinfeng}\ \bibnamefont {Liao}},\ }\bibfield  {title}
  {\enquote {\bibinfo {title} {{Quantifying the chiral magnetic effect from
  anomalous-viscous fluid dynamics}},}\ }\href {\doibase
  10.1088/1674-1137/42/1/011001} {\bibfield  {journal} {\bibinfo  {journal}
  {Chin. Phys. C}\ }\textbf {\bibinfo {volume} {42}},\ \bibinfo {pages}
  {011001} (\bibinfo {year} {2018})},\ \Eprint
  {http://arxiv.org/abs/1611.04586} {arXiv:1611.04586 [nucl-th]} \BibitemShut
  {NoStop}%
\bibitem [{\citenamefont {An}\ \emph {et~al.}(2022)\citenamefont {An} \emph
  {et~al.}}]{An:2021wof}%
  \BibitemOpen
  \bibfield  {author} {\bibinfo {author} {\bibfnamefont {Xin}\ \bibnamefont
  {An}} \emph {et~al.},\ }\bibfield  {title} {\enquote {\bibinfo {title} {{The
  BEST framework for the search for the QCD critical point and the chiral
  magnetic effect}},}\ }\href {\doibase 10.1016/j.nuclphysa.2021.122343}
  {\bibfield  {journal} {\bibinfo  {journal} {Nucl. Phys. A}\ }\textbf
  {\bibinfo {volume} {1017}},\ \bibinfo {pages} {122343} (\bibinfo {year}
  {2022})},\ \Eprint {http://arxiv.org/abs/2108.13867} {arXiv:2108.13867
  [nucl-th]} \BibitemShut {NoStop}%
\bibitem [{\citenamefont {Shen}\ \emph {et~al.}(2016)\citenamefont {Shen},
  \citenamefont {Qiu}, \citenamefont {Song}, \citenamefont {Bernhard},
  \citenamefont {Bass},\ and\ \citenamefont {Heinz}}]{Shen:2014vra}%
  \BibitemOpen
  \bibfield  {author} {\bibinfo {author} {\bibfnamefont {Chun}\ \bibnamefont
  {Shen}}, \bibinfo {author} {\bibfnamefont {Zhi}\ \bibnamefont {Qiu}},
  \bibinfo {author} {\bibfnamefont {Huichao}\ \bibnamefont {Song}}, \bibinfo
  {author} {\bibfnamefont {Jonah}\ \bibnamefont {Bernhard}}, \bibinfo {author}
  {\bibfnamefont {Steffen}\ \bibnamefont {Bass}}, \ and\ \bibinfo {author}
  {\bibfnamefont {Ulrich}\ \bibnamefont {Heinz}},\ }\bibfield  {title}
  {\enquote {\bibinfo {title} {{The iEBE-VISHNU code package for relativistic
  heavy-ion collisions}},}\ }\href {\doibase 10.1016/j.cpc.2015.08.039}
  {\bibfield  {journal} {\bibinfo  {journal} {Comput. Phys. Commun.}\ }\textbf
  {\bibinfo {volume} {199}},\ \bibinfo {pages} {61--85} (\bibinfo {year}
  {2016})},\ \Eprint {http://arxiv.org/abs/1409.8164} {arXiv:1409.8164
  [nucl-th]} \BibitemShut {NoStop}%
\bibitem [{\citenamefont {Bloczynski}\ \emph {et~al.}(2013)\citenamefont
  {Bloczynski}, \citenamefont {Huang}, \citenamefont {Zhang},\ and\
  \citenamefont {Liao}}]{Bloczynski:2012en}%
  \BibitemOpen
  \bibfield  {author} {\bibinfo {author} {\bibfnamefont {John}\ \bibnamefont
  {Bloczynski}}, \bibinfo {author} {\bibfnamefont {Xu-Guang}\ \bibnamefont
  {Huang}}, \bibinfo {author} {\bibfnamefont {Xilin}\ \bibnamefont {Zhang}}, \
  and\ \bibinfo {author} {\bibfnamefont {Jinfeng}\ \bibnamefont {Liao}},\
  }\bibfield  {title} {\enquote {\bibinfo {title} {{Azimuthally fluctuating
  magnetic field and its impacts on observables in heavy-ion collisions}},}\
  }\href {\doibase 10.1016/j.physletb.2012.12.030} {\bibfield  {journal}
  {\bibinfo  {journal} {Phys. Lett. B}\ }\textbf {\bibinfo {volume} {718}},\
  \bibinfo {pages} {1529--1535} (\bibinfo {year} {2013})},\ \Eprint
  {http://arxiv.org/abs/1209.6594} {arXiv:1209.6594 [nucl-th]} \BibitemShut
  {NoStop}%
\bibitem [{\citenamefont {Xu}\ \emph {et~al.}(2018{\natexlab{a}})\citenamefont
  {Xu}, \citenamefont {Zhao}, \citenamefont {Wang}, \citenamefont {Li},
  \citenamefont {Lin}, \citenamefont {Shen},\ and\ \citenamefont
  {Wang}}]{Xu:2017qfs}%
  \BibitemOpen
  \bibfield  {author} {\bibinfo {author} {\bibfnamefont {Hao-jie}\ \bibnamefont
  {Xu}}, \bibinfo {author} {\bibfnamefont {Jie}\ \bibnamefont {Zhao}}, \bibinfo
  {author} {\bibfnamefont {Xiaobao}\ \bibnamefont {Wang}}, \bibinfo {author}
  {\bibfnamefont {Hanlin}\ \bibnamefont {Li}}, \bibinfo {author} {\bibfnamefont
  {Zi-Wei}\ \bibnamefont {Lin}}, \bibinfo {author} {\bibfnamefont {Caiwan}\
  \bibnamefont {Shen}}, \ and\ \bibinfo {author} {\bibfnamefont {Fuqiang}\
  \bibnamefont {Wang}},\ }\bibfield  {title} {\enquote {\bibinfo {title}
  {{Varying the chiral magnetic effect relative to flow in a single
  nucleus-nucleus collision}},}\ }\href {\doibase
  10.1088/1674-1137/42/8/084103} {\bibfield  {journal} {\bibinfo  {journal}
  {Chin. Phys. C}\ }\textbf {\bibinfo {volume} {42}},\ \bibinfo {pages}
  {084103} (\bibinfo {year} {2018}{\natexlab{a}})},\ \Eprint
  {http://arxiv.org/abs/1710.07265} {arXiv:1710.07265 [nucl-th]} \BibitemShut
  {NoStop}%
\bibitem [{\citenamefont {Voloshin}(2018)}]{Voloshin:2018qsm}%
  \BibitemOpen
  \bibfield  {author} {\bibinfo {author} {\bibfnamefont {Sergei~A.}\
  \bibnamefont {Voloshin}},\ }\bibfield  {title} {\enquote {\bibinfo {title}
  {{Estimate of the signal from the chiral magnetic effect in heavy-ion
  collisions from measurements relative to the participant and spectator flow
  planes}},}\ }\href {\doibase 10.1103/PhysRevC.98.054911} {\bibfield
  {journal} {\bibinfo  {journal} {Phys. Rev. C}\ }\textbf {\bibinfo {volume}
  {98}},\ \bibinfo {pages} {054911} (\bibinfo {year} {2018})},\ \Eprint
  {http://arxiv.org/abs/1805.05300} {arXiv:1805.05300 [nucl-ex]} \BibitemShut
  {NoStop}%
\bibitem [{\citenamefont {Zhao}\ \emph {et~al.}(2019)\citenamefont {Zhao},
  \citenamefont {Li},\ and\ \citenamefont {Wang}}]{Zhao:2017nfq}%
  \BibitemOpen
  \bibfield  {author} {\bibinfo {author} {\bibfnamefont {Jie}\ \bibnamefont
  {Zhao}}, \bibinfo {author} {\bibfnamefont {Hanlin}\ \bibnamefont {Li}}, \
  and\ \bibinfo {author} {\bibfnamefont {Fuqiang}\ \bibnamefont {Wang}},\
  }\bibfield  {title} {\enquote {\bibinfo {title} {{Isolating the chiral
  magnetic effect from backgrounds by pair invariant mass}},}\ }\href {\doibase
  10.1140/epjc/s10052-019-6671-1} {\bibfield  {journal} {\bibinfo  {journal}
  {Eur. Phys. J. C}\ }\textbf {\bibinfo {volume} {79}},\ \bibinfo {pages} {168}
  (\bibinfo {year} {2019})},\ \Eprint {http://arxiv.org/abs/1705.05410}
  {arXiv:1705.05410 [nucl-ex]} \BibitemShut {NoStop}%
\bibitem [{\citenamefont {Magdy}\ \emph {et~al.}(2018)\citenamefont {Magdy},
  \citenamefont {Shi}, \citenamefont {Liao}, \citenamefont {Ajitanand},\ and\
  \citenamefont {Lacey}}]{Magdy:2017yje}%
  \BibitemOpen
  \bibfield  {author} {\bibinfo {author} {\bibfnamefont {Niseem}\ \bibnamefont
  {Magdy}}, \bibinfo {author} {\bibfnamefont {Shuzhe}\ \bibnamefont {Shi}},
  \bibinfo {author} {\bibfnamefont {Jinfeng}\ \bibnamefont {Liao}}, \bibinfo
  {author} {\bibfnamefont {N.}~\bibnamefont {Ajitanand}}, \ and\ \bibinfo
  {author} {\bibfnamefont {Roy~A.}\ \bibnamefont {Lacey}},\ }\bibfield  {title}
  {\enquote {\bibinfo {title} {{New correlator to detect and characterize the
  chiral magnetic effect}},}\ }\href {\doibase 10.1103/PhysRevC.97.061901}
  {\bibfield  {journal} {\bibinfo  {journal} {Phys. Rev. C}\ }\textbf {\bibinfo
  {volume} {97}},\ \bibinfo {pages} {061901} (\bibinfo {year} {2018})},\
  \Eprint {http://arxiv.org/abs/1710.01717} {arXiv:1710.01717
  [physics.data-an]} \BibitemShut {NoStop}%
\bibitem [{\citenamefont {Tang}(2020)}]{Tang:2019pbl}%
  \BibitemOpen
  \bibfield  {author} {\bibinfo {author} {\bibfnamefont {A.~H.}\ \bibnamefont
  {Tang}},\ }\bibfield  {title} {\enquote {\bibinfo {title} {{Probe chiral
  magnetic effect with signed balance function}},}\ }\href {\doibase
  10.1088/1674-1137/44/5/054101} {\bibfield  {journal} {\bibinfo  {journal}
  {Chin. Phys. C}\ }\textbf {\bibinfo {volume} {44}},\ \bibinfo {pages}
  {054101} (\bibinfo {year} {2020})},\ \Eprint
  {http://arxiv.org/abs/1903.04622} {arXiv:1903.04622 [nucl-ex]} \BibitemShut
  {NoStop}%
\bibitem [{\citenamefont {Wen}\ \emph {et~al.}(2018)\citenamefont {Wen},
  \citenamefont {Bryon}, \citenamefont {Wen},\ and\ \citenamefont
  {Wang}}]{Wen:2016zic}%
  \BibitemOpen
  \bibfield  {author} {\bibinfo {author} {\bibfnamefont {Fufang}\ \bibnamefont
  {Wen}}, \bibinfo {author} {\bibfnamefont {Jacob}\ \bibnamefont {Bryon}},
  \bibinfo {author} {\bibfnamefont {Liwen}\ \bibnamefont {Wen}}, \ and\
  \bibinfo {author} {\bibfnamefont {Gang}\ \bibnamefont {Wang}},\ }\bibfield
  {title} {\enquote {\bibinfo {title} {{Event-shape-engineering study of charge
  separation in heavy-ion collisions}},}\ }\href {\doibase
  10.1088/1674-1137/42/1/014001} {\bibfield  {journal} {\bibinfo  {journal}
  {Chin. Phys. C}\ }\textbf {\bibinfo {volume} {42}},\ \bibinfo {pages}
  {014001} (\bibinfo {year} {2018})},\ \Eprint
  {http://arxiv.org/abs/1608.03205} {arXiv:1608.03205 [nucl-th]} \BibitemShut
  {NoStop}%
\bibitem [{\citenamefont {Zhang}\ and\ \citenamefont
  {Jia}(2022)}]{Zhang:2021kxj}%
  \BibitemOpen
  \bibfield  {author} {\bibinfo {author} {\bibfnamefont {Chunjian}\
  \bibnamefont {Zhang}}\ and\ \bibinfo {author} {\bibfnamefont {Jiangyong}\
  \bibnamefont {Jia}},\ }\bibfield  {title} {\enquote {\bibinfo {title}
  {{Evidence of Quadrupole and Octupole Deformations in Zr96+Zr96 and Ru96+Ru96
  Collisions at Ultrarelativistic Energies}},}\ }\href {\doibase
  10.1103/PhysRevLett.128.022301} {\bibfield  {journal} {\bibinfo  {journal}
  {Phys. Rev. Lett.}\ }\textbf {\bibinfo {volume} {128}},\ \bibinfo {pages}
  {022301} (\bibinfo {year} {2022})},\ \Eprint
  {http://arxiv.org/abs/2109.01631} {arXiv:2109.01631 [nucl-th]} \BibitemShut
  {NoStop}%
\bibitem [{\citenamefont {Xu}\ \emph {et~al.}(2021)\citenamefont {Xu},
  \citenamefont {Li}, \citenamefont {Wang}, \citenamefont {Shen},\ and\
  \citenamefont {Wang}}]{Xu:2021vpn}%
  \BibitemOpen
  \bibfield  {author} {\bibinfo {author} {\bibfnamefont {Hao-jie}\ \bibnamefont
  {Xu}}, \bibinfo {author} {\bibfnamefont {Hanlin}\ \bibnamefont {Li}},
  \bibinfo {author} {\bibfnamefont {Xiaobao}\ \bibnamefont {Wang}}, \bibinfo
  {author} {\bibfnamefont {Caiwan}\ \bibnamefont {Shen}}, \ and\ \bibinfo
  {author} {\bibfnamefont {Fuqiang}\ \bibnamefont {Wang}},\ }\bibfield  {title}
  {\enquote {\bibinfo {title} {{Determine the neutron skin type by relativistic
  isobaric collisions}},}\ }\href {\doibase 10.1016/j.physletb.2021.136453}
  {\bibfield  {journal} {\bibinfo  {journal} {Phys. Lett. B}\ }\textbf
  {\bibinfo {volume} {819}},\ \bibinfo {pages} {136453} (\bibinfo {year}
  {2021})},\ \Eprint {http://arxiv.org/abs/2103.05595} {arXiv:2103.05595
  [nucl-th]} \BibitemShut {NoStop}%
\bibitem [{\citenamefont {Shi}\ \emph {et~al.}(2019)\citenamefont {Shi},
  \citenamefont {Zhang}, \citenamefont {Hou},\ and\ \citenamefont
  {Liao}}]{Shi:2018sah}%
  \BibitemOpen
  \bibfield  {author} {\bibinfo {author} {\bibfnamefont {Shuzhe}\ \bibnamefont
  {Shi}}, \bibinfo {author} {\bibfnamefont {Hui}\ \bibnamefont {Zhang}},
  \bibinfo {author} {\bibfnamefont {Defu}\ \bibnamefont {Hou}}, \ and\ \bibinfo
  {author} {\bibfnamefont {Jinfeng}\ \bibnamefont {Liao}},\ }\bibfield  {title}
  {\enquote {\bibinfo {title} {{Chiral Magnetic Effect in Isobaric Collisions
  from Anomalous-Viscous Fluid Dynamics (AVFD)}},}\ }\href {\doibase
  10.1016/j.nuclphysa.2018.10.007} {\bibfield  {journal} {\bibinfo  {journal}
  {Nucl. Phys. A}\ }\textbf {\bibinfo {volume} {982}},\ \bibinfo {pages}
  {539--542} (\bibinfo {year} {2019})},\ \Eprint
  {http://arxiv.org/abs/1807.05604} {arXiv:1807.05604 [hep-ph]} \BibitemShut
  {NoStop}%
\bibitem [{\citenamefont {Xu}\ \emph {et~al.}(2018{\natexlab{b}})\citenamefont
  {Xu}, \citenamefont {Wang}, \citenamefont {Li}, \citenamefont {Zhao},
  \citenamefont {Lin}, \citenamefont {Shen},\ and\ \citenamefont
  {Wang}}]{Xu:2017zcn}%
  \BibitemOpen
  \bibfield  {author} {\bibinfo {author} {\bibfnamefont {Hao-Jie}\ \bibnamefont
  {Xu}}, \bibinfo {author} {\bibfnamefont {Xiaobao}\ \bibnamefont {Wang}},
  \bibinfo {author} {\bibfnamefont {Hanlin}\ \bibnamefont {Li}}, \bibinfo
  {author} {\bibfnamefont {Jie}\ \bibnamefont {Zhao}}, \bibinfo {author}
  {\bibfnamefont {Zi-Wei}\ \bibnamefont {Lin}}, \bibinfo {author}
  {\bibfnamefont {Caiwan}\ \bibnamefont {Shen}}, \ and\ \bibinfo {author}
  {\bibfnamefont {Fuqiang}\ \bibnamefont {Wang}},\ }\bibfield  {title}
  {\enquote {\bibinfo {title} {{Importance of isobar density distributions on
  the chiral magnetic effect search}},}\ }\href {\doibase
  10.1103/PhysRevLett.121.022301} {\bibfield  {journal} {\bibinfo  {journal}
  {Phys. Rev. Lett.}\ }\textbf {\bibinfo {volume} {121}},\ \bibinfo {pages}
  {022301} (\bibinfo {year} {2018}{\natexlab{b}})},\ \Eprint
  {http://arxiv.org/abs/1710.03086} {arXiv:1710.03086 [nucl-th]} \BibitemShut
  {NoStop}%
\bibitem [{\citenamefont {Hammelmann}\ \emph {et~al.}(2020)\citenamefont
  {Hammelmann}, \citenamefont {Soto-Ontoso}, \citenamefont {Alvioli},
  \citenamefont {Elfner},\ and\ \citenamefont {Strikman}}]{Hammelmann:2019vwd}%
  \BibitemOpen
  \bibfield  {author} {\bibinfo {author} {\bibfnamefont {Jan}\ \bibnamefont
  {Hammelmann}}, \bibinfo {author} {\bibfnamefont {Alba}\ \bibnamefont
  {Soto-Ontoso}}, \bibinfo {author} {\bibfnamefont {Massimiliano}\ \bibnamefont
  {Alvioli}}, \bibinfo {author} {\bibfnamefont {Hannah}\ \bibnamefont
  {Elfner}}, \ and\ \bibinfo {author} {\bibfnamefont {Mark}\ \bibnamefont
  {Strikman}},\ }\bibfield  {title} {\enquote {\bibinfo {title} {{Influence of
  the neutron-skin effect on nuclear isobar collisions at energies available at
  the BNL Relativistic Heavy Ion Collider}},}\ }\href {\doibase
  10.1103/PhysRevC.101.061901} {\bibfield  {journal} {\bibinfo  {journal}
  {Phys. Rev. C}\ }\textbf {\bibinfo {volume} {101}},\ \bibinfo {pages}
  {061901} (\bibinfo {year} {2020})},\ \Eprint
  {http://arxiv.org/abs/1908.10231} {arXiv:1908.10231 [nucl-th]} \BibitemShut
  {NoStop}%
\bibitem [{\citenamefont {{STAR Collaboration}}(2022)}]{hepdata.115993}%
  \BibitemOpen
  \bibfield  {author} {\bibinfo {author} {\bibnamefont {{STAR
  Collaboration}}},\ }\href@noop {} {\enquote {\bibinfo {title} {{Search for
  the chiral magnetic effect with isobar collisions at $\sqrt {s_{NN}}$=200 GeV
  by the STAR Collaboration at the BNL Relativistic Heavy Ion Collider}},}\
  }\bibinfo {howpublished} {{HEPData (collection)}} (\bibinfo {year} {2022}),\
  \bibinfo {note} {\url{https://doi.org/10.17182/hepdata.115993}}\BibitemShut
  {NoStop}%
\bibitem [{\citenamefont {Feng}\ \emph {et~al.}(2022)\citenamefont {Feng},
  \citenamefont {Zhao}, \citenamefont {Li}, \citenamefont {Xu},\ and\
  \citenamefont {Wang}}]{Feng:2021pgf}%
  \BibitemOpen
  \bibfield  {author} {\bibinfo {author} {\bibfnamefont {Yicheng}\ \bibnamefont
  {Feng}}, \bibinfo {author} {\bibfnamefont {Jie}\ \bibnamefont {Zhao}},
  \bibinfo {author} {\bibfnamefont {Hanlin}\ \bibnamefont {Li}}, \bibinfo
  {author} {\bibfnamefont {Hao-jie}\ \bibnamefont {Xu}}, \ and\ \bibinfo
  {author} {\bibfnamefont {Fuqiang}\ \bibnamefont {Wang}},\ }\bibfield  {title}
  {\enquote {\bibinfo {title} {{Two- and three-particle nonflow contributions
  to the chiral magnetic effect measurement by spectator and participant planes
  in relativistic heavy ion collisions}},}\ }\href {\doibase
  10.1103/PhysRevC.105.024913} {\bibfield  {journal} {\bibinfo  {journal}
  {Phys. Rev. C}\ }\textbf {\bibinfo {volume} {105}},\ \bibinfo {pages}
  {024913} (\bibinfo {year} {2022})},\ \Eprint
  {http://arxiv.org/abs/2106.15595} {arXiv:2106.15595 [nucl-ex]} \BibitemShut
  {NoStop}%
\bibitem [{\citenamefont {Nijs}\ and\ \citenamefont {van~der
  Schee}(2021)}]{Nijs:2021kvn}%
  \BibitemOpen
  \bibfield  {author} {\bibinfo {author} {\bibfnamefont {Govert}\ \bibnamefont
  {Nijs}}\ and\ \bibinfo {author} {\bibfnamefont {Wilke}\ \bibnamefont {van~der
  Schee}},\ }\bibfield  {title} {\enquote {\bibinfo {title} {{Inferring nuclear
  structure from heavy isobar collisions using Trajectum}},}\ }\href@noop {} {\
   (\bibinfo {year} {2021})},\ \Eprint {http://arxiv.org/abs/2112.13771}
  {arXiv:2112.13771 [nucl-th]} \BibitemShut {NoStop}%
\bibitem [{Sup()}]{SuppMat}%
  \BibitemOpen
  \href@noop {} {}\bibinfo {note} {See Supplemental Material at
  \url{https://journals.aps.org/prc/supplemental/10.1103/PhysRevC.106.L051903}
  for technical details on the simulations and analysis that have been used to
  obtain the results reported in the main manuscript.}\BibitemShut {Stop}%
\bibitem [{\citenamefont {Milton}\ \emph {et~al.}(2021)\citenamefont {Milton},
  \citenamefont {Wang}, \citenamefont {Sergeeva}, \citenamefont {Shi},
  \citenamefont {Liao},\ and\ \citenamefont {Huang}}]{Milton:2021wku}%
  \BibitemOpen
  \bibfield  {author} {\bibinfo {author} {\bibfnamefont {Ryan}\ \bibnamefont
  {Milton}}, \bibinfo {author} {\bibfnamefont {Gang}\ \bibnamefont {Wang}},
  \bibinfo {author} {\bibfnamefont {Maria}\ \bibnamefont {Sergeeva}}, \bibinfo
  {author} {\bibfnamefont {Shuzhe}\ \bibnamefont {Shi}}, \bibinfo {author}
  {\bibfnamefont {Jinfeng}\ \bibnamefont {Liao}}, \ and\ \bibinfo {author}
  {\bibfnamefont {Huan~Zhong}\ \bibnamefont {Huang}},\ }\bibfield  {title}
  {\enquote {\bibinfo {title} {{Utilization of event shape in search of the
  chiral magnetic effect in heavy-ion collisions}},}\ }\href {\doibase
  10.1103/PhysRevC.104.064906} {\bibfield  {journal} {\bibinfo  {journal}
  {Phys. Rev. C}\ }\textbf {\bibinfo {volume} {104}},\ \bibinfo {pages}
  {064906} (\bibinfo {year} {2021})},\ \Eprint
  {http://arxiv.org/abs/2110.01435} {arXiv:2110.01435 [nucl-th]} \BibitemShut
  {NoStop}%
\bibitem [{\citenamefont {Lacey}\ \emph {et~al.}(2022)\citenamefont {Lacey},
  \citenamefont {Magdy}, \citenamefont {Parfenov},\ and\ \citenamefont
  {Taranenko}}]{Lacey:2022baf}%
  \BibitemOpen
  \bibfield  {author} {\bibinfo {author} {\bibfnamefont {Roy~A.}\ \bibnamefont
  {Lacey}}, \bibinfo {author} {\bibfnamefont {Niseem}\ \bibnamefont {Magdy}},
  \bibinfo {author} {\bibfnamefont {Petr}\ \bibnamefont {Parfenov}}, \ and\
  \bibinfo {author} {\bibfnamefont {Arkadiy}\ \bibnamefont {Taranenko}},\
  }\bibfield  {title} {\enquote {\bibinfo {title} {{Scaling properties of
  background- and chiral-magnetically-driven charge separation in heavy ion
  collisions at $\sqrt s_{\mathrm{NN}}=200$ GeV}},}\ }\href@noop {} {\
  (\bibinfo {year} {2022})},\ \Eprint {http://arxiv.org/abs/2203.10029}
  {arXiv:2203.10029 [nucl-ex]} \BibitemShut {NoStop}%
\bibitem [{\citenamefont {Feng}\ \emph {et~al.}(2021)\citenamefont {Feng},
  \citenamefont {Lin}, \citenamefont {Zhao},\ and\ \citenamefont
  {Wang}}]{Feng:2021oub}%
  \BibitemOpen
  \bibfield  {author} {\bibinfo {author} {\bibfnamefont {Yicheng}\ \bibnamefont
  {Feng}}, \bibinfo {author} {\bibfnamefont {Yufu}\ \bibnamefont {Lin}},
  \bibinfo {author} {\bibfnamefont {Jie}\ \bibnamefont {Zhao}}, \ and\ \bibinfo
  {author} {\bibfnamefont {Fuqiang}\ \bibnamefont {Wang}},\ }\bibfield  {title}
  {\enquote {\bibinfo {title} {{Revisit the chiral magnetic effect expectation
  in isobaric collisions at the relativistic heavy ion collider}},}\ }\href
  {\doibase 10.1016/j.physletb.2021.136549} {\bibfield  {journal} {\bibinfo
  {journal} {Phys. Lett. B}\ }\textbf {\bibinfo {volume} {820}},\ \bibinfo
  {pages} {136549} (\bibinfo {year} {2021})},\ \Eprint
  {http://arxiv.org/abs/2103.10378} {arXiv:2103.10378 [nucl-ex]} \BibitemShut
  {NoStop}%
\bibitem [{\citenamefont {Abdallah}\ \emph
  {et~al.}(2022{\natexlab{b}})\citenamefont {Abdallah} \emph
  {et~al.}}]{STAR:2021pwb}%
  \BibitemOpen
  \bibfield  {author} {\bibinfo {author} {\bibfnamefont {Mohamed}\ \bibnamefont
  {Abdallah}} \emph {et~al.} (\bibinfo {collaboration} {STAR}),\ }\bibfield
  {title} {\enquote {\bibinfo {title} {{Search for the Chiral Magnetic Effect
  via Charge-Dependent Azimuthal Correlations Relative to Spectator and
  Participant Planes in Au+Au Collisions at $\sqrt{s_{NN}}$ =\, 200\,GeV}},}\
  }\href {\doibase 10.1103/PhysRevLett.128.092301} {\bibfield  {journal}
  {\bibinfo  {journal} {Phys. Rev. Lett.}\ }\textbf {\bibinfo {volume} {128}},\
  \bibinfo {pages} {092301} (\bibinfo {year} {2022}{\natexlab{b}})},\ \Eprint
  {http://arxiv.org/abs/2106.09243} {arXiv:2106.09243 [nucl-ex]} \BibitemShut
  {NoStop}%
\end{thebibliography}%

\end{document}